\documentclass[a4paper,11pt]{article}
\pdfoutput=1 % if your are submitting a pdflatex (i.e. if you have
\usepackage{jheppub} % for details on the use of the package, please
\usepackage{main-defs}

\usepackage[T1]{fontenc} % if needed
\usepackage{graphicx}
\usepackage{placeins}
\usepackage{xspace}
\usepackage{soul}
\usepackage{siunitx}
\usepackage{upgreek}
\usepackage{subcaption}

\usepackage[dvipsnames]{xcolor}
\def\be{\begin{equation}}
\def\ee{\end{equation}}
\def\bea{\begin{eqnarray}}%https://www.overleaf.com/project/5ff2cabfd2991befdc91c143
\def\eea{\end{eqnarray}}

\preprint{\begin{flushright} DESY 21-019 \\ 
IFIC/21-03 \end{flushright}}

\title{\boldmath Hunting wino and higgsino dark matter at the muon collider with disappearing tracks}

\author[a,b]{Rodolfo Capdevilla,}
\author[c]{Federico Meloni,}
\author[d]{Rosa Simoniello,}
\author[e]{Jose Zurita}

\affiliation[a]{Department of Physics, University of Toronto, Canada}
\affiliation[b]{Perimeter Institute for Theoretical Physics, Waterloo, Ontario, Canada}
\affiliation[c]{Deutsches Elektronen-Synchrotron DESY, Hamburg, Germany}
\affiliation[d]{CERN, Geneva, Switzerland}
\affiliation[e]{Instituto de F\'{\i}sica Corpuscular, CSIC-Universitat de Val\`encia, Valencia, Spain}

\emailAdd{rcapdevilla@perimeterinstitute.ca}
\emailAdd{federico.meloni@desy.de}
\emailAdd{rosa.simoniello@cern.ch}
\emailAdd{jzurita@ific.uv.es}

\abstract{We study the capabilities of a muon collider experiment to detect disappearing tracks originating when a heavy and electrically charged long-lived particle decays via $X^+ \to Y^+ Z^0$, where $X^+$ and $Z^0$ are two almost mass degenerate new states and $Y^+$ is a charged Standard Model particle.
The backgrounds induced by the in-flight decays of the muon beams (BIB) can create detector hit combinations that mimic long-lived particle signatures, making the search a daunting task. We design a simple strategy to tame the BIB, based on a detector-hit-level selection exploiting timing information and hit-to-hit correlations, followed by simple requirements on the quality of reconstructed tracks. Our strategy allows us to reduce the number of tracks from BIB to an average of 0.08 per event, hence being able to design a cut-and-count analysis that shows that it is possible to cover weak doublets and triplets with masses close to $\sqrt{s}/2$ in the 0.1-10 ns range. In particular, this implies that a 10 TeV muon collider is able to probe thermal MSSM higgsinos and thermal MSSM winos, thus rivaling the FCC-hh in that respect, and further enlarging the physics program of the muon collider into the territory of WIMP dark matter and long-lived signatures. We also provide parton-to-reconstructed level efficiency maps, allowing an estimation of the coverage of disappearing tracks at muon colliders for arbitrary models.
} 

\begin{document} 
\maketitle
\flushbottom

\section{Introduction}
\label{sec:intro}

Long-Lived Particles (LLPs) have become an important and active area of research in the last few years. LLPs appear in a variety of models trying to address fundamental puzzles of the Standard Model (SM) of particle physics (for a comprehensive review of the theoretical foundations see e.g. Ref.~\cite{Curtin:2018mvb}), yield a large palette of signatures at colliders (for a review see e.g. Ref.~\cite{Alimena:2019zri}), and furthermore inspired the design of dedicated, small-scale detectors, such as MATHUSLA~\cite{Chou:2016lxi}, CODEX-b~\cite{Gligorov:2017nwh} and FASER~\cite{Feng:2017uoz}.
The latter, with its fast approval and construction, is a shining example of the important role that LLP searches are taking within the Beyond Standard Model (BSM) searches at the LHC.

The realm of LLP signatures at colliders can be broadly classified into three distinct classes that depend upon the LLP quantum numbers under the SM gauge group. We will refer to them as \emph{neutral}, \emph{dark showers}, and \emph{charged}, respectively\footnote{Milli-charged particles might yield long-lived signatures, but their phenomenology is analogous to the neutral case, hence we will not consider them further.}. The first class is relatively unconstrained, since neutral particles travel unscathed through detector material. Their presence can only be inferred by their decay products (within jargon sometimes this is referred to as \emph{indirect detection} of LLPs), and the coverage of the LHC and previous colliders strongly depends on the LLP lifetime, which has an upper limit of about $c \tau \sim 10^{7}$ m  stemming from Big Bang nucleosynthesis considerations~\cite{Jedamzik:2006xz}. These neutral LLPs arise in the context of so called ``portal'' models, and can also be relatively light (down to sub-GeV), hence they have been thoroughly studied in the context of the ``Physics Beyond Collider" initiative, see~\cite{Beacham:2019nyx}. The second class originates from strongly interacting confining dark sectors, as is the case in e.g. Hidden Valley Scenarios~\cite{Strassler:2006im}. In analogy to quantum chromodynamics, the dark quarks shower and hadronise in the dark sector, and eventually decay to SM final states, leaving a myriad of displaced vertices in the detector (emerging jets~\cite{Schwaller:2015gea}, semi-visible jets~\cite{Cohen:2015toa}, dark jets~\cite{Park:2017rfb}, etc). While these LLP are indeed SM neutral, it is their high multiplicity that merits a separate category. Last, but not least, the charged LLPs have strong constraints from previous colliders as LEP and Tevatron, as it would be impossible to avoid these particles coupling to the electroweak gauge bosons with typical weak coupling strength. Hence in this category we necessarily would have $m_{LLP} \gtrsim 100$~GeV, and, depending on their lifetime, the collider signatures would differ considerably.

This article focuses on disappearing tracks (DT), a class of these LLP charged signatures, consisting of a charged particle travelling for few centimeters or less, hence manifesting itself as an ``incomplete'' track with missing hits in the outermost layers of the tracking system, which we will refer to as a ``tracklet'' in the following. In addition, associated to this tracklet, there is little to no energy deposit in the calorimeters and no hits in the muon system. Over recent years, DT has proven to be a powerful tool to probe models that predict the existence of particles with lifetime of $\mathcal{O}(ns)$ (see e.g. Refs.~\cite{Belyaev:2016lok,Khoze:2017ixx,Mahbubani:2017gjh,Fukuda:2017jmk,Lopez-Honorez:2017ora,Calibbi:2018fqf,Saito:2019rtg,Bharucha:2018pfu,Belanger:2018sti,Filimonova:2018qdc,Jana:2019tdm,Chiang:2020rcv,Belyaev:2020wok,Calibbi:2021fld}) and is actively looked for at the LHC by the ATLAS and CMS collaborations~\cite{Aaboud:2017mpt,Aad:2013yna,Sirunyan:2020pjd,CMS:2014gxa}. Among the many theoretical scenarios that give rise to long-lived charged particles, an important motivation comes from dark matter (DM). The lack of a DM candidate in the SM is one of the most compelling arguments to seek for its extension. Among the various possibilities, Weakly Interacting Massive Particles (WIMP) are a well-motivated option to obtain the correct relic abundance through the freeze-out mechanism. Higgsino- and wino-like states, the supersymmetric partners of the Higgs and W fields respectively, are a notable model-specific example of a WIMP. Minimalistic bottom-up SM extensions \cite{Cirelli:2005uq} propose to add new multiplets to the SM such that the lightest neutral component is stable and provides a DM candidate. Depending on the mass hierarchy and differences between the particles in the multiplets, striking experimental signatures can be obtained, with either charged states with a long enough lifetime to be observed directly as charged tracks, or in the production of DTs. The cases of {\it pure} higgsino and wino DM in the MSSM are the {\it de-facto} benchmark not only of the LHC collaborations, but also of phenomenological work, see e.g. Refs.~\cite{Delgado:2019tbz,McKay:2017xlc,Chun:2016cnm,Jung:2015boa,Fukuda:2017jmk,Mahbubani:2017gjh,Chakraborti:2017dpu,Bramante:2015una,Giudice:1998xp,Randall:1998uk}.

The coverage for wino and higgsino DM at the LHC falls well below the mass values where the relic density constraint saturates i.e. 2.7 and 1.1 TeV, respectively~\cite{Hisano:2006nn}. This coverage could have been much larger had the LHC been optimised for LLP signals in its original design, although in general, the investigation of either wino or higgsino DM scenarios is particularly challenging at hadron colliders, see e.g.~\cite{Low:2014cba,Cirelli:2014dsa,diCortona:2016fsn,Mahbubani:2017gjh,Chigusa:2019zae,Saito:2019rtg}. Future lepton colliders, such as a high-energy muon collider (MuC) could greatly extend the reach of the current hadronic machines (LHC and HL-LHC) \cite{Delahaye:2019omf,Long:2020wfp}. A multi-TeV MuC has numerous physics motivations that make it particularly appealing. It has been proven that a MuC is effectively a high-luminosity vector boson collider \cite{Costantini:2020stv}. With this, and the fact that vector boson fusion (VBF) processes grow with energy, one can use a MuC to study Higgs couplings \cite{Chiesa:2020awd,Han:2020pif,Buttazzo:2020uzc}. It has also been shown that a MuC can discover the new physics behind the $(g-2)_\mu$ anomaly \cite{Capdevilla:2020qel,Buttazzo:2020eyl,Yin:2020afe,Capdevilla:2021rwo}, assuming that the current running experiments at Fermilab \cite{Fienberg:2019ddu} and JPARC \cite{Sato:2017/P} establish the $(g-2)_\mu$ excess as a source of new physics. In addition to these motivations, the reach of a multi-TeV MuC has recently been studied in the context of a few BSM scenarios~\cite{Huang:2021nkl,Liu:2021jyc,Chen:2021rnl,Han:2021udl}. Here we add another important motivation, namely we show that such a MuC can be able to cover thermal DM~\cite{Han:2020uak}  beyond the capabilities of its main competitor, a putative high energy proton-proton collider \cite{Arkani-Hamed:2015vfh,Strategy:2019vxc}. In this article we perform the first realistic assessment of the sensitivity to DTs of the proposed MuC, using the widespread higgsino and wino benchmarks.

While the common lore is that lepton colliders provide a ``clean'' experimental environment, at a MuC this is not entirely correct. The products of the in-flight decays of the muon beams and the results of their interactions with the detector and beamline material, usually known as ``Beam Induced Backgrounds'' (BIB), create a large particle flux that interacts with the detector elements. This in turn can cause the reconstruction of spurious DT candidates from the large number of detector hits. Because of this effect, the BIB is the primary source of background for the disappearing track signal at the MuC. In this article we perform the first realistic assessment of this background using full detector simulation of the most up-to-date MuC detector design. We find that the huge BIB can be tamed by exploiting timing information, quality criteria on the tracks and the fact that the signal tends to be centrally produced, while the BIB is largely in the forward direction. This careful assessment of the BIB is an important result of our analysis.

The rest of this article is structured as follows. In Section~\ref{sec:theory} we discuss the phenomenology of charged LLPs at colliders, and then we briefly review the MSSM pure higgsino and pure wino benchmarks. In Section~\ref{sec:detector} we present our working setup for the MuC, and discuss in detail the event generation of the samples and the parameterisation of the detector response, including the use of the simulation software \GEANT~4 for the track detector. In Section~\ref{sec:bib} we study in detail the BIB, showing how it can be efficiently suppressed by a series of selection criteria that render it negligible. As a by-product, we derive the tracklet reconstruction efficiency at the MuC in a model independent manner. Finally, in Section~\ref{sec:analysis} we describe our analysis strategy and show that the MuC can cover the pure higgsino and pure wino scenarios, with a reach competitive to that of a high-energy hadron collider. We defer our conclusions to Section~\ref{sec:conclusions}.

\section{Long lived charged particles at colliders}
\label{sec:theory}

We start by considering the potential collider signatures of $X^{\pm} \to Z^0 Y^{\pm}$, which depend upon the mass ratios $y=m_{Y} / m_{X}$ and $z= m_{Z}/ m_{X}$. The experimental signature also depends upon the mean proper decay length of $X$ (in comparison with the position of the different detector components), which we would refer to as $c \tau$.
In this description, the WIMP DM case we consider corresponds to the $y \to 0, z \to 1$ limit.
Since these are charged particles, necessarily $m_X \gtrsim 100$ GeV, with the actual bound depending on the model details. The possible signatures for a charged LLP are, then\footnote{We note that charged particles with electric charges different from one would also lead to anomalous ionisation that can be exploited. Such signatures are important for the search for e.g. magnetic monopoles. Large ionisation could offer an additional handle in the search for heavy WIMPs, but we will ignore these cases in what follows.}: 
\begin{itemize}
\item $X^{\pm}$ is reconstructed as a track that makes it past the inner detector: \emph{Heavy Stable Charged Particle, HSCP}, or \emph{out-of-time decays} ($c \tau \gtrsim 1 m$). The HSCP may be heavily ionising. $Y^{\pm}$ is likely to escape detection.
\item $X^{\pm}$ is reconstructed as a tracklet (DT signature). It is relevant for decay lengths $c \tau <1$~m, namely, for the inner tracking detectors. Such experimental signatures lose coverage for very low lifetimes, where the $X^{\pm}$ does not traverse enough detector layers to be reconstructed.
\item $Y^{\pm}$ is reconstructed as a displaced track, ignoring the decay vertex (displaced $Y^{\pm}$, where the case of $Y^{\pm}=\pi^{\pm}$ was carried out in Ref.~\cite{Curtin:2017bxr} for e-p colliders).
\item $X^{\pm}$ and $Y^{\pm}$ are \emph{both} reconstructed and connected in a \emph{kinked track}.
\end{itemize}
The presence of the undetectable $Z^0$ can lead to sizeable momentum imbalance in the event if the $X^{\pm}$ decay occurs within the detector.

In this article we focus on the disappearing track signature, taking MSSM {\it pure} electroweakinos as our case study. We anyhow stress that our results are general and widely applicable to more general classes of models. We note that all other signatures, with the exception of the HSCP, are extremely challenging at the LHC. Hence these constitute an ideal target for a lepton-lepton or lepton-hadron collider.

\begin{figure}[h]
\begin{center}
    \begin{subfigure}[t]{0.48\textwidth}
    	\centering
		\includegraphics[width=\textwidth]{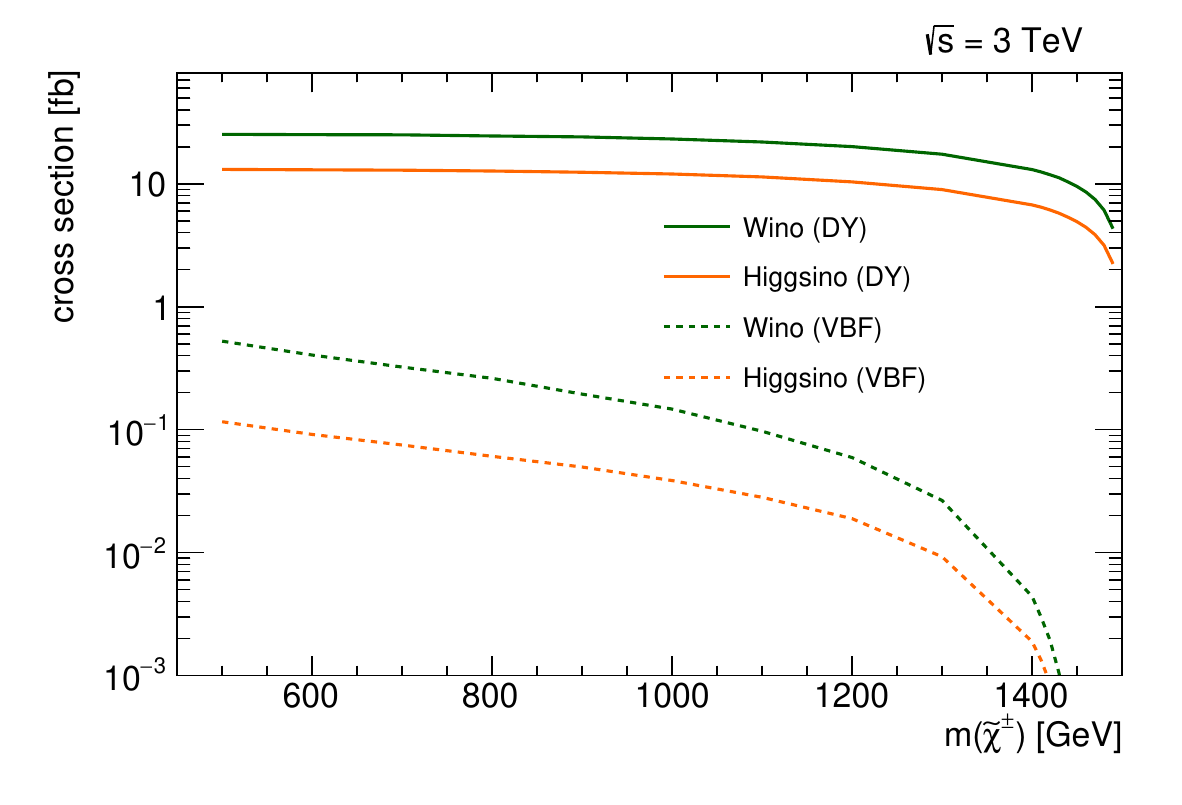}
%		\caption{}
	\end{subfigure}%
    \begin{subfigure}[t]{0.48\textwidth}
    	\centering
    	\includegraphics[width=\textwidth]{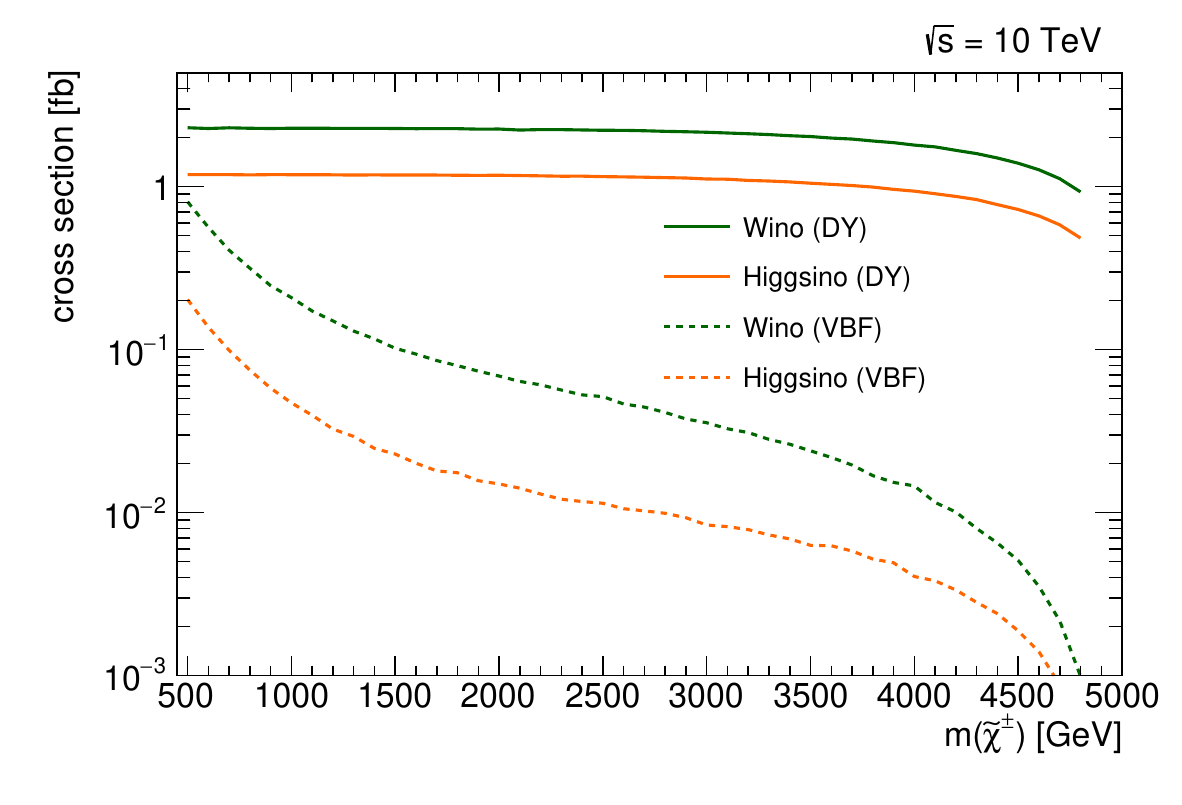}
%		\caption{}
	\end{subfigure}
\end{center}
\caption{Production cross sections for $\chipm\chimp$ at $\sqrt{s}=3 (10)$ TeV, in the left (right) panel. We show the cases of pure wino (green) and pure higgsino (orange). The full (dashed) lines corresponds to Drell-Yan (VBF) production.}
\label{fig:muc_chargino_xs}
\end{figure}

From the perspective of dark matter, models where DM is the lightest neutral component of a single electroweak multiplet with non-trivial $SU(2)_L$ charges (which qualify as WIMPs) would feature at least one charged particle, whose mass is split from the neutral one by a few hundred MeV, owing to radiative corrections from electroweak gauge bosons~\cite{Thomas:1998wy}.
Hence, these sub-GeV mass splittings in the dark sector are very natural, and that compressed spectrum implies that the charged particle can have a macroscopically appreciable decay length\footnote{Note that this is not the only mechanism to have long-lived charged particles in DM models, they could also arise if DM is produced by e.g. freeze-in from parent decay~\cite{Belanger:2018sti,Calibbi:2021fld}.}.
MSSM electroweakinos have been widely reviewed in the literature, see e.g. Ref.~\cite{Canepa:2020ntc} and references therein. For the purpose of the current work, it suffices to state the main features of the so-called ``pure'' cases, namely where in the low energy spectrum only the wino ($\tilde{W}$) or higgsino ($\tilde{H}$) are kept and all other supersymmetric states are decoupled. The low energy spectrum features one charged particle, the chargino $\chipm$, and one (two) neutral particle(s) for the wino (higgsino) scenario. These particles are collectively referred to as electroweakinos, and are odd under a discrete $\mathbf{Z}_2$ symmetry, known as R-parity, which stabilises the lightest neutral particle rendering it an excellent DM candidate. 
Direct detection of that DM occurs via Higgs exchange (as by construction the SM Z-boson current is either null or strongly inhibited) which is suppressed by mixing effects\footnote{The reason for this suppression is that the Higgs connects a higgsino with a bino or wino, hence in the ``pure cases" one of the electroweakinos is decoupled.}. Due to one-loop radiative corrections, after the breaking of $SU(2)_L$, the charged state splits from the neutral one by 166 (344) MeV for wino (higgsino), giving rise to a mean proper decay length of 6 cm (6.6 mm) for the relic favoured mass. The production of pairs of electroweakinos at a MuC proceeds mainly via an s-channel photon or off-shell Z-boson (Drell-Yan production, or DY), with other processes, such as vector boson fusion (VBF), being subdominant. 
We present in  Figure~\ref{fig:muc_chargino_xs} the production cross sections for pure wino and pure higgsino electroweakino pairs, as a function of the electroweakino masses, for centre of mass energies of 3 TeV (left) and 10 TeV (right), considering production from Drell-Yan and also (for completeness) from VBF. 
The details of the event generation are described in Section~\ref{sec:sim}. From this figure we see that both a $\sqrt{s}=3$ and a $\sqrt{s}=10$ TeV MuC would produce electroweakino pairs with appreciable rates for both pure wino and pure higgsino scenarios, provided they are kinematically accessible.
Note that while the 3~TeV collider has approximately one order of magnitude higher cross section than its 10~TeV counterpart, the final number of expected events is similar. This is the case because it is envisioned that the 10~TeV collider will collect about 10~times more integrated luminosity than the 3~TeV one~\cite{Delahaye:2019omf,Long:2020wfp}. Both colliders expect to produce about 10000 electroweakino pairs via DY, while VBF production would yield only a handful of events.

\begin{figure}[h]
\begin{center}
    \begin{subfigure}[t]{0.48\textwidth}
    	\centering
		\includegraphics[width=\textwidth]{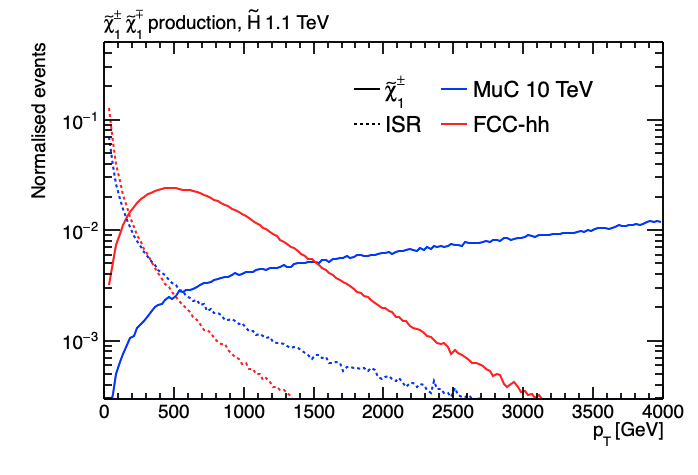}
		\caption{}
	\end{subfigure}%
    \begin{subfigure}[t]{0.48\textwidth}
    	\centering
    	\includegraphics[width=\textwidth]{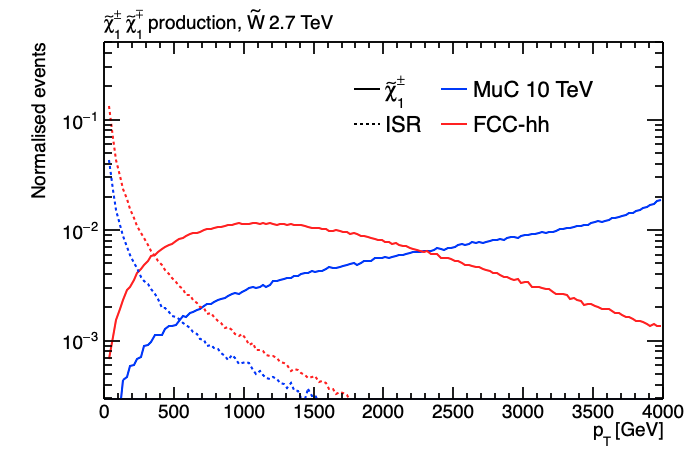}
    	\caption{}
	\end{subfigure}\\
    \begin{subfigure}[t]{0.48\textwidth}
    	\centering
		\includegraphics[width=\textwidth]{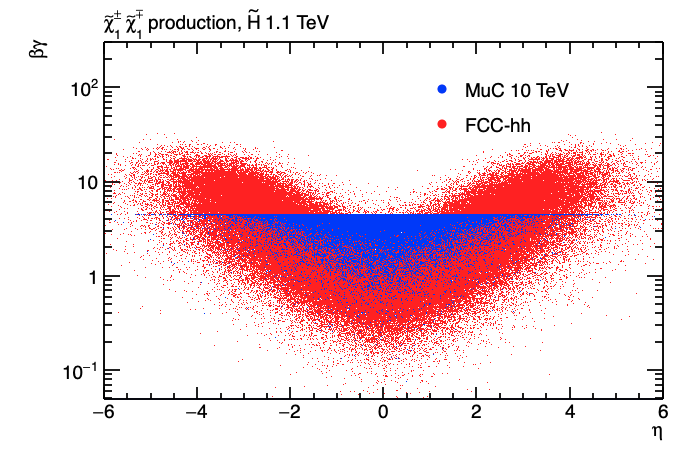}
		\caption{}
	\end{subfigure}%
    \begin{subfigure}[t]{0.48\textwidth}
    	\centering
    	\includegraphics[width=\textwidth]{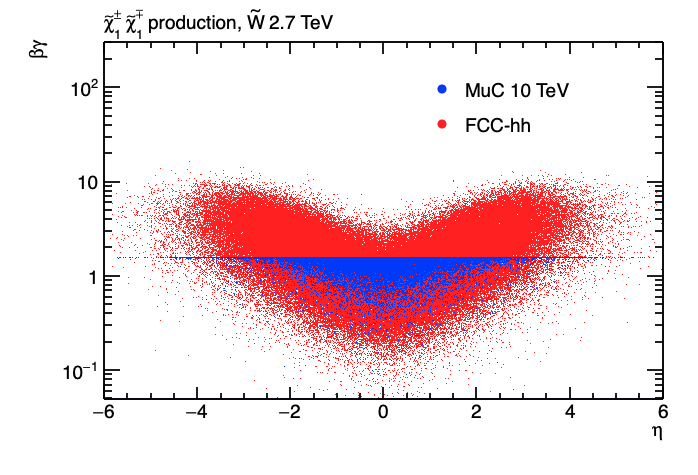}
    	\caption{}
	\end{subfigure}%
\end{center}
\caption{Kinematic distributions for thermal higgsino, $m_\chi$=1.1 TeV (left) and thermal wino, $m_\chi = $2.7 TeV  (right) at a putative Muon Collider with $\sqrt{s}=10$ TeV and at a proton-proton collider with $\sqrt{s}=100$ TeV. The upper panel shows the \pT of the chargino (solid line) and of the corresponding initial state radiation particle (jet for FCC-hh, photon for MuC) with dashed lines. The lower panel shows the distribution of the events in the $\eta- \beta \gamma$ plane, using red (blue) for the FCC-hh (MuC).}
\label{fig:fcc_vs_muc_distros}
\end{figure}

As an appetiser to our main result, we present kinematical distributions for a MuC with $\sqrt{s} = 10$ TeV and the FCC-hh ($pp, \sqrt{s}=100$ TeV) in  Figure~\ref{fig:fcc_vs_muc_distros}, for thermal higgsinos (left) and winos (right). 
In the upper panel we plot the \pT of the \chipm, and of the corresponding additional radiation (jet for FCC-hh, photon for the MuC). From the upper panel we immediately see that the chargino has a harder \pT at the MuC than at the FCC-hh. This is an encouraging feature, as higher momentum tracks are less likely to be mimicked by the backgrounds. In the lower panel we see that while the FCC-hh enjoys a large longitudinal boost, the events at the MuC are far more central. This observation supports having a detector with a reduced polar angle acceptance, which is in line with the current plans for the MuC detector design (see Section~\ref{sec:detector}). Indeed, selecting parton level tracklets satisfying $|\eta| < 2.44$ keeps 98 (97.5) \% of the thermal wino (higgsino) events. 
In addition, we also see that the maximum Lorentz factor $\beta \gamma$ is smaller at the MuC than at the FCC by about one order of magnitude. However the overall distribution has a less significant spread to lower values and the MuC is expected to efficiently detect charged tracks with lower $c \tau$ than the FCC-hh, as we will discuss in Section~\ref{sec:detector}.

We note that in principle pure wino and higgsino could be probed by indirect detection at e.g. the Cherenkov Telescope Array (CTA)~\cite{Hryczuk:2019nql,Rinchiuso:2020skh} or the AMS-02 experiment~\cite{Krall:2017xij} before the timescale of the next generation of colliders (FCC, MuC, etc). A potential excess from these indirect detection experiments would not invalidate the collider program, but rather give more strength to it, as a direct probe of the wino and higgsino (including a characterisation of the dark sector particle properties) under controlled conditions would become a necessity. 

\section{A high-energy muon collider}
\label{sec:detector}

The main experimental facility considered in this study is a future muon collider able to operate at centre of mass energies of $\sqrt{s} = 3$ and 10~TeV. For the $\sqrt{s} = 3$~TeV (10~TeV) configuration we assume that an integrated luminosity of 1~(10)~ab$^{-1}$ will be collected. 

\begin{figure}[h]
\begin{center}
    \centering
	\includegraphics[width=0.85\textwidth]{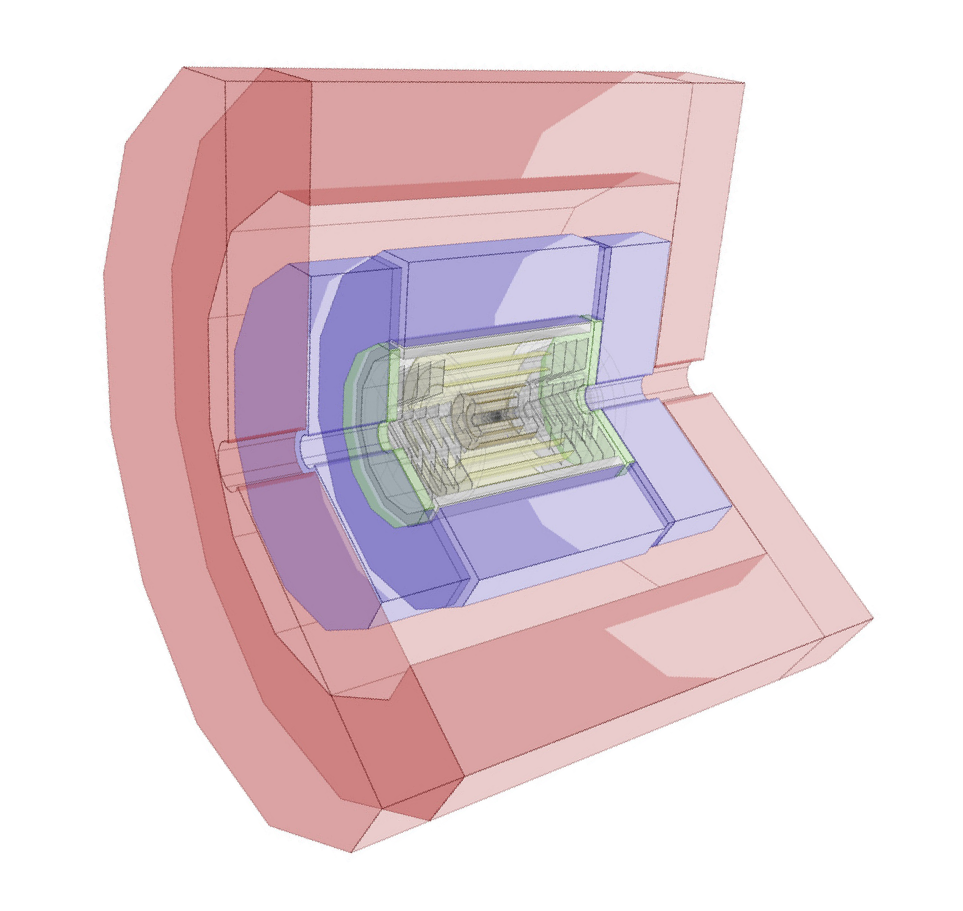}
\end{center}
\caption{Illustration of the full detector, from the \GEANT 4 model. Different colours represent different sub-detector systems: the innermost region, highlighted in the yellow shade, represents the tracking detectors. The green and purple elements represent the calorimeter system, while the red outermost shell represents the magnet return yoke instrumented with muon chambers. The space between the calorimeters and the return yoke will be occupied by a 3.57~T solenoid magnet (not displayed).}
\label{fig:detector}
\end{figure}

The detector model is based on the SiD ILC concept~\cite{behnke2007ilc} with few changes, described in the following. The detector, shown in  Figure~\ref{fig:detector}, is designed with a cylindrical layout\footnote{A right-handed coordinate system with its origin at the nominal interaction point in the centre of the detector is used. Cylindrical coordinates $(r,\phi)$ are used in the transverse plane, $\phi$ being the azimuthal angle around the z-axis and $\theta$ the polar angle with respect to the z-axis. The pseudorapidity is defined in terms of the polar angle $\theta$ as $\eta = -\ln \tan(\theta/2)$.}. The innermost system consists of a full-silicon tracking detector. The tracking detector is surrounded by a calorimeter system, and is immersed in a solenoidal magnetic field of 3.57~T. Finally, the outermost part of the detector consists of a magnet yoke designed to contain the return flux of the magnetic field and is instrumented with muon chambers. 

The studies presented in this work focus on the tracking detector: the vertex detector (VXD) is the innermost tracking system. It consists of four concentric cylindrical shells equipped with double-layers of position-sensitive detectors (distanced 2~mm) in the barrel region extending up to $z=\pm~6.5$~cm and positioned at radii of 3.1, 5.1, 7.4, 10.2~cm; and four double-layer (distanced 2~mm) endcap disks on each detector side with radius of 11.2~cm and positioned at a $z$ of 8, 12, 20, 28~cm. The VXD envisions the use of squared silicon pixel sensors with a single point resolution of \SI{5}{\micro\meter} and a time resolution of 30~ps. 
A particularly relevant quantity when comparing different collider options is the minimum radial distance for a tracklet to be efficiently detected. For example, the current ATLAS pixel detector is able to reconstruct tracks down to 12~cm\footnote{In all these values we are considering the position of the 4th hit as the minimal radial distance for a tracklet to be reconstructed.}, due to the Insertable B-Layer upgrade~\cite{Capeans:1291633,CERN-LHCC-2012-009,Abbott:2018ikt}.
However, the HL-LHC inner detector would move this number to 22~cm~\cite{CERN-LHCC-2017-021}, and existing FCC-hh designs contemplate 10-15 cm~\cite{Saito:2019rtg}, albeit with $|\eta| < 2.3-2.6$, which results in a lower signal efficiency (cf. Figure~\ref{fig:fcc_vs_muc_distros}). In contrast, CLIC and the MuC are expected to detect tracks with $c \tau = 10.2$ cm.

\begin{figure}[h]
\begin{center}
    \begin{subfigure}[t]{0.48\textwidth}
    	\centering
		\includegraphics[width=\textwidth]{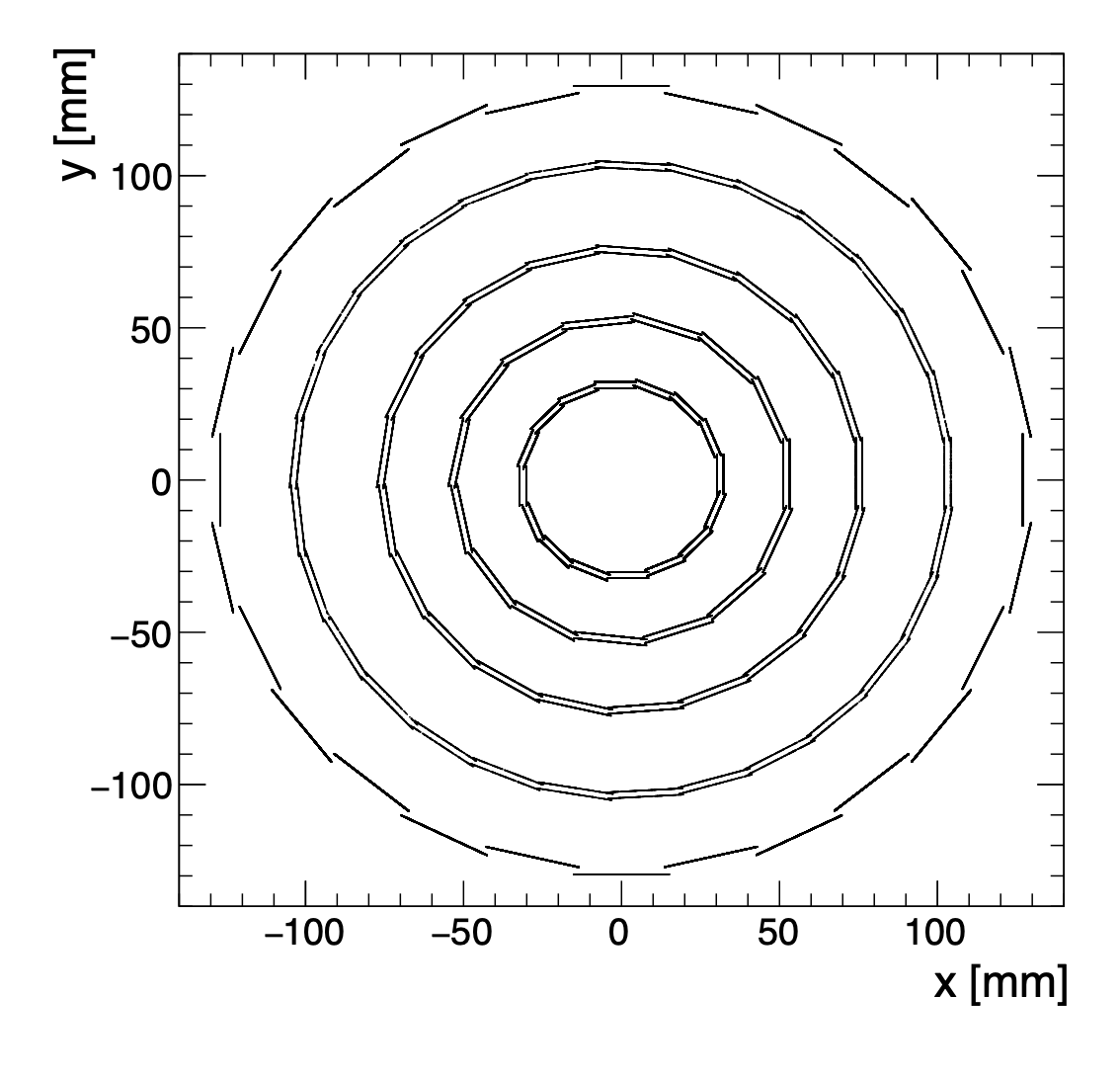}
		\caption{}
	\end{subfigure}%
    \begin{subfigure}[t]{0.48\textwidth}
    	\centering
    	\includegraphics[width=\textwidth]{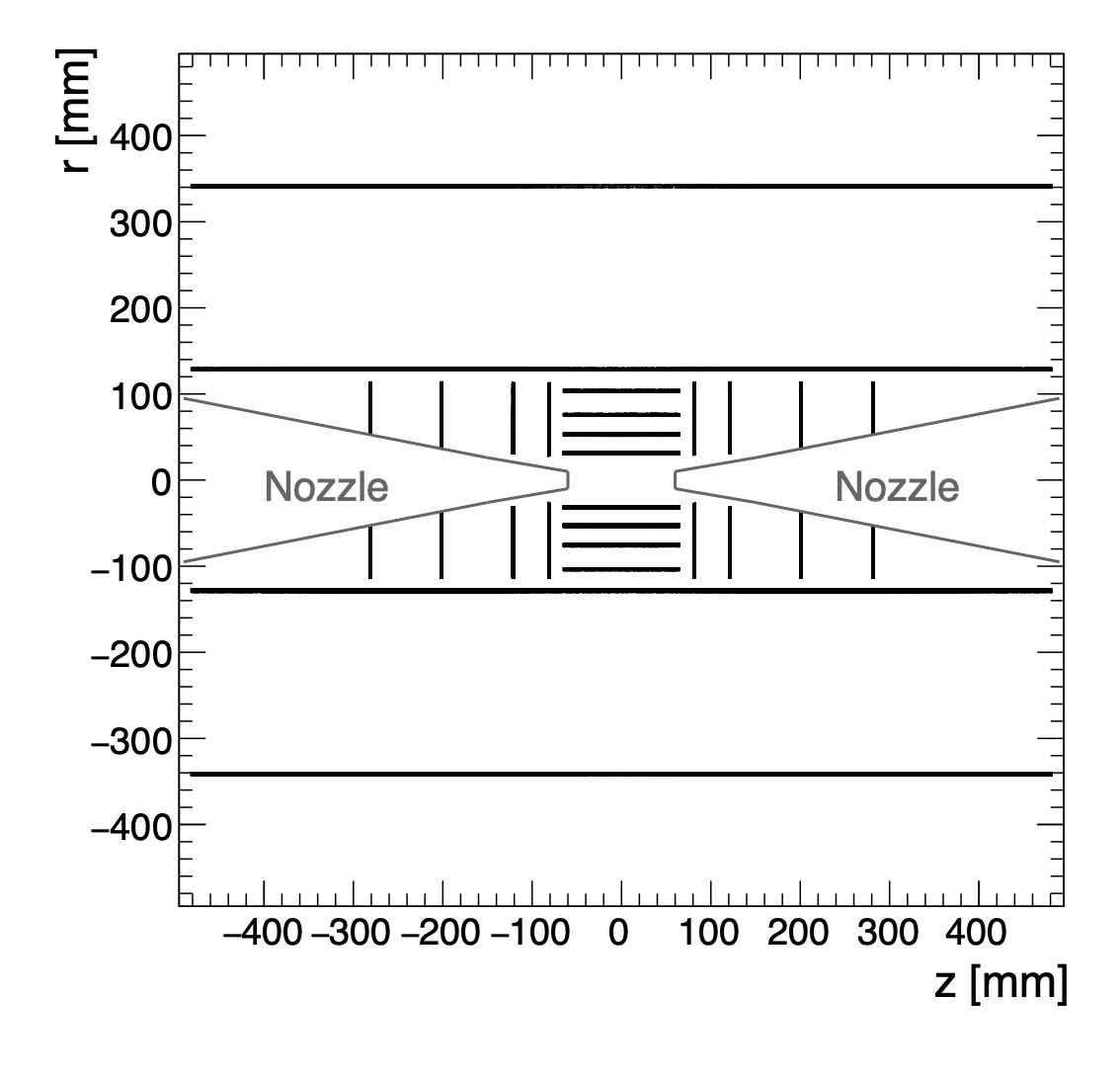}
		\caption{}
	\end{subfigure}
\end{center}
\caption{Transverse (a) and longitudinal (b) views of the innermost region of the tracking detectors. In the left panel, one can visualise the four concentric cylindrical double-layers of the VXD and the innermost layer of the IT. These appear also in the right panel along with the four double-layer VXD endcap disks and the second layer of the IT.}
\label{fig:tracker}
\end{figure}

Around the vertex detector lies the tracker detector, which is divided in an inner (IT) and outer part. The first inner barrel layer is used in this analysis to detect tracks that disappeared as will be discussed in Section~\ref{subsec:tracklets}. It consists of a silicon single-layer of radius 12.7~cm covering up to $z = \pm~48.16$~cm, with a single point resolution ($R\phi \times z$) of \SI{7}{}$\times$\SI{90}{\micro\meter} and a time resolution of 60~ps\footnote{The vertex and tracker detector geometries have not been fully optimised yet for a muon collider and they may change in the future. We do not expect a large impact on the feasibility of this analysis, but the final design will determine the lifetime coverage.}. 
 Figure~\ref{fig:tracker} shows a detailed view in the transverse and longitudinal planes of the vertex and inner tracker detectors in the region that will be used in this analysis. In order to reduce the amount of BIB particles entering the interaction region, two tungsten shielding cones (“nozzles”) are placed in the forward regions along the beam axis at $|\eta| > 2.44$.

\subsection{Simulated event samples and detector parameterisation}
\label{sec:sim}

The detector response is modelled with a hybrid approach. The response of the tracking detectors, crucial for this analysis, is modelled using a detector simulation~\cite{frank_markus_2018_1464634} based on \GEANT 4~\cite{Agostinelli:2002hh}. For the remaining detector systems, we employ response functions for the high-level objects reconstruction and identification efficiencies using the Delphes~\cite{deFavereau:2013fsa} program. The simulated events used to derive the tracking detector response functions were overlaid with BIB events simulated with the MARS15 software~\cite{Mokhov:2017klc}.

The simulation of the BIB is a crucial element to assess the power of the presented search. The muon decay products and the products of their interaction with the machine elements can reach the interaction region and the detectors. The BIB simulation has been performed for machines with a centre of mass energy of $\sqrt{s} = 1.5$~TeV and $\sqrt{s} = 125$~GeV~\cite{Mokhov:2011zzd,Mokhov:2014hza,Bartosik:2019dzq,Lucchesi:2020dku,DiBenedetto:2018cpy}. The composition, flux, and energy spectra of the BIB surviving the shielding and entering the detector depend on the machine configuration and collision energy. The most important BIB property is that it is composed of low-energy particles. For $\sqrt{s} = 1.5$~TeV collisions the BIB mostly consists of $\mathcal{O}(1)$~MeV photons and electrons and $\mathcal{O}(100)$~MeV hadrons; and is characterised by a broad arrival time in the detector. The particle and hit multiplicity was observed to mildly decrease with increasing centre of mass energy~\cite{Bartosik:2019dzq}. 
Increasing the centre of mass energy of the MuC has two main effects on the BIB: the particle and hit multiplicity mildly decreases and the measured energy deposits get distributed at larger rapidities~\cite{Mokhov:2017klc}. While the energy deposition in the detector by the BIB could increase at higher collision energies, the analysis presented in this work is only sensitive to these two main effects. On one hand, the hit multiplicity affects tracking reconstruction. On the other hand, the angular distribution of the tracks helps to discriminate signal over background, as we will discuss below, the signal tracks will be more central whereas BIB tracks will be mostly forward. For these reasons, in the absence of a dedicated BIB simulation at the centre of mass energies used in this study (certainly a desirable feature), the simulated events with $\sqrt{s} = 1.5$~TeV were taken as a conservative estimate of the BIB. We also note that the rate of BIB particles arriving in the detector can be optimised in the design of the shielding nozzles. While this could result in a decreased angular acceptance, the work described in this paper is not expected to be affected as only events in the central region of the detector are considered.

Monte Carlo samples were used to predict the expected backgrounds from SM processes and to model the signal scenarios under consideration. Signal and background processes were generated with \textsc{MadGraph5}\_aMC@NLO~2.8.2~\cite{Alwall:2014hca} interfaced to \textsc{Pythia}~8.244 \cite{Sjostrand:2014zea} for the parton showering and hadronisation. The matrix element calculation was performed at tree level and includes the emission of up to one additional photon for all the relevant samples. 

Two classes of signal events were generated assuming either pure wino or pure higgsino scenarios. In both scenarios, events were generated including $\chipm\chimp$ pair production. The \chipm mass is varied for the generation from 500~GeV to $\sqrt{s}/2$, with the \ninoone (and \ninotwo) masses set according to the splittings defined in Section~\ref{sec:theory}. While in the pure higgsino and pure wino scenarios the lifetime is a function of the mass only (varying by less than 10\% for the scanned range), we consider here the lifetime as a free parameter, as customary in LLP searches at the LHC. 
As discussed in section \ref{sec:theory}, only DY and charged VBF production processes were included in the generation.
Production processes initiated by the photons coming from the collinear radiation of the high-energy muon beams are neglected and considered sub-dominant~\cite{Han:2020uak}. The production cross sections were taken from the generator prediction at leading-order, and were reported in  Figure~\ref{fig:muc_chargino_xs}. For example, a thermal wino, $m_{\chipm} = 2.7$~TeV, has a cross section of 2.2~fb for DY production and 0.039~fb for VBF at a MuC operating at $\sqrt{s} = 10$~TeV; while a thermal higgsino, $m_{\chipm} = 1.1$~TeV, has a cross section of 11.4~fb for DY production and 0.028~fb for VBF at a MuC operating at $\sqrt{s} = 3$~TeV

Assuming that the background arising from SM particles that are reconstructed as tracklets can be reduced to a negligible contribution as discussed in Section~\ref{subsec:selection}, the most significant SM backgrounds arise from the $\mu\mu \rightarrow \nu\nu$ process, for which we used a generator prediction cross section of about 55~pb (3~pb if requiring the presence of an additional photon with $p_T >$ 10 GeV), roughly independent of the $\sqrt{s}$ in the range considered in this work. Other SM backgrounds with final states including hadrons or charged leptons are neglected, as they are expected to have a negligible impact on the analysis after the event selection. The contribution of other SM processes with sizeable cross sections where all visible particles would be produced out of the detector acceptance has been checked and found to be negligible. 

\subsection{Reconstruction}
\label{sec:reco}

The reconstruction of physics objects is performed at truth-level with parameterised detector response functions~\cite{deFavereau:2013fsa}. In particular, we derived dedicated parameterisations for the reconstructed tracklet momentum, its reconstruction efficiency and fake rate. For the extraction of these dedicated tracking response functions in the presence of BIB, track reconstruction is performed on the simulated samples using the \textit{conformal tracking}~\cite{Brondolin:2019awm} package within the muon collider reconstruction software suite~\cite{Frank_2014,frank_markus_2018_1464634,Frank_2015,GAEDE2006177,MuCSW} that we optimised for the reconstruction of tracklets. A comprehensive description of the method is presented in Ref.~\cite{Brondolin:2019awm}, and a short summary including the configuration used for this analysis is reported in this section. 

The conformal track finding algorithm is based on a cellular automaton \cite{GLAZOV1993262,KISEL200685} algorithm running in the conformal plane $(u,v)$. 
The conformal plane is defined by dividing the coordinates of each hit in the transverse plane $(x,y)$ by the squared radial distance, $r^2=x^2+y^2$. This transforms the circular trajectories in the transverse plane into linear trajectories, and it has advantage to speed up the track finding. 
Some deviations from the linear behaviour are expected due to the multiple scattering of the particles in the detector material and non-prompt tracks. These effects can be accounted in the tuning of the conformal track finding algorithm; nevertheless, they are not relevant for this analysis where the aim is to reconstruct prompt tracks with extremely high \pt, as shown in  Figure~\ref{fig:fcc_vs_muc_distros}. 

The cellular automaton algorithm uses a set of local criteria such as the distance and the angular difference in the conformal plane between two hits in consecutive detector layers to create connections. A connection between two hits is called a cell. Cells are accepted if the angle in the conformal plane between the two hits is less than a parameter $\Theta_{\mathrm{max}}$ and if the cell length is less than $l_{\mathrm{max}}$. Cells can be connected, lengthening the track candidate, if the angle with the starting cell is less than a parameter $\alpha_{\mathrm{max}}$. 
For each cell, a weight variable indicates the number of consecutive connections made. Track candidates are obtained by following the connected cells from highest to lowest weight. If two or more track candidates share more than one hit, the track with the larger number of hits is preferred, if the tracks have the same number of hits the one with the best $\chi^2$ from the track fit is taken. 
Tracks are accepted if they have a number of hits equal or greater than $N_{\mathrm{min}}^{\mathrm{hits}}$, and if their $\chi^2$ is less than $\chi_{\mathrm{max}}^2$. The variables 
$\Theta_{\mathrm{max}}$, $l_{\mathrm{max}}$, $\alpha_{\mathrm{max}}$, $N_{\mathrm{min}}^{\mathrm{hits}}$, $\chi_{\mathrm{max}}^2$ are all tunable parameters of the algorithm. 

%The cellular automaton algorithm uses a set of local criteria such as the distance and the angular difference ($\alpha$ and $\Theta$ in the conformal and longitudinal plane, respectively) between two hits in consecutive detector layers to create connections. A connection between two hits is called a cell. For each cell, a weight variable indicates the number of consecutive connections made. Track candidates are obtained by following the connected cells from highest to lowest weight. If two or more track candidates share more than one hit, the track with the larger number of hits is preferred, if the tracks have the same number of hits the one with the best $\chi^2$ from the track fit is taken. 

The reconstruction algorithm starts by running on a given collection of seeding hits and can then be used to extend the track candidates to subsequent hit collections or to re-run on the unused hits with different selection thresholds to recover more challenging tracks. The configuration with the relevant parameters used in this analysis is reported in Table~\ref{tab:tracking-config}.

\begin{table}[!htbp]
\begin{center}
\setlength{\tabcolsep}{11pt}
{\small
\begin{tabular*}{\textwidth}{cccccccc}
\hline
Step & Function & Hit collection & 
$\alpha_{\mathrm{max}}$ & 
$\Theta_{\mathrm{max}}$ & 
$\chi_{\mathrm{max}}^2$ & 
$N_{\mathrm{min}}^{\mathrm{hits}}$ &
$l_{\mathrm{max}} $  \\ 
 &  &  & [rad] & [rad] &  &  & [mm$^{-1}$] \\ 
\hline
1 & Seeding & Vertex Barrel & 0.005&	0.005 &	100	& 4 &	0.020  \\
\hline
2 & Seeding & Vertex Barrel + & 0.007&	0.007 &	100	& 4 &	0.020  \\
 &  & 2 Endcap Layers  & & &	& &   \\
\hline
3 & Extension & $1^{\mathrm{st}}$ Tracker Layer & 0.050 &	0.050 &	100	& 6 &	0.009  \\
\hline
\end{tabular*}
%%%
}
\end{center}
\caption{Configuration of the conformal tracking used in this analysis. 
%The parameters $\alpha_{\mathrm{max}}$ and $\Theta_{\mathrm{max}}$ are the upper thresholds on the angle between two hits in the conformal and longitudinal plane, respectively, 
The parameters $\Theta_{\mathrm{max}}$ and $\alpha_{\mathrm{max}}$ are the upper thresholds on the angle to connect two hits or two cells, respectively, 
whereas $\chi_{\mathrm{max}}^2$ is the upper threshold on both linear regression fits performed in the conformal and longitudinal planes. The minimum number of hits required in a track is indicated by $N_{\mathrm{min}}^{\mathrm{hits}}$, while the upper threshold on the cell length in the conformal plane is referred to as $l_{\mathrm{max}}$.}
\label{tab:tracking-config}
\end{table}

The direction of the track finding is chosen from innermost to outermost layers and is composed of three steps. The algorithm starts by seeding tracks in the vertex barrel detector satisfying tight criteria. Then the track building is performed with loosened criteria and the hits of the first two double layers of endcaps of the vertex detector are included to increase the signal acceptance up to $\theta\sim \pi/6$. No attempt at reconstructing tracks in the region of $\theta < \pi/6$ or $\theta > 5/6\pi$  was done because of the overwhelming contribution of the beam induced background in this region and the central nature of the signal. Finally, the track candidates are extended to the first barrel layer of the inner tracker. Given that larger distances in the cartesian plane are equivalent to smaller distances in the conformal plane (since cartesian coordinates are divided by the squared radius), the threshold on the $l_{\mathrm{max}}$ parameter is reduced in this step.

Because of the large number of hit combinations from the BIB, regional track finding in six orthogonal sectors of the polar angle was used to reduce the time needed for the reconstruction process. Future approaches may provide a better scaling with the large number of hits and resulting combinations and eliminate the need for regional track finding.

As a last step, the track candidates are fitted to estimate the track parameters using a Kalman filter~\cite{FRUHWIRTH1987444}.

\FloatBarrier
\section{Beam-induced background rejection}
\label{sec:bib}

There are no SM processes that manifest themselves with a disappearing track experimental signature.
The experimental sources of disappearing tracks are either catastrophic interactions of a charged particle with the detector material, or arise from random combinations of hits from uncorrelated particles. The large hit multiplicity in the tracking detectors due to the presence of the BIB make the potential to reconstruct such spurious ``fake'' tracklets from the accidental alignment of hits the main source of background.

Our strategy to reject these ``fake'' tracklets is based on a two-step process. First, we apply selection criteria on the detector hits that are given as inputs to the track reconstruction. The second step rejects track candidates based on a set of quality criteria on the fitted tracks.  
We would like to stress that our BIB rejection strategy is fairly general and could also be applied for other LLP studies that predict charged particles originating from the interaction region, or that have a small displacement (up to a few centimeters). The detailed strategy is explained in the following sections.

\subsection{Hit-level rejection}
\label{subsec:hit}

The arrival time and direction of the particles from the BIB can be exploited to substantially reduce the background contamination in the collection of hits used for the track reconstruction.

\begin{figure}[h]
\begin{center}
    \begin{subfigure}[t]{0.495\textwidth}
    	\centering
		\includegraphics[width=\textwidth]{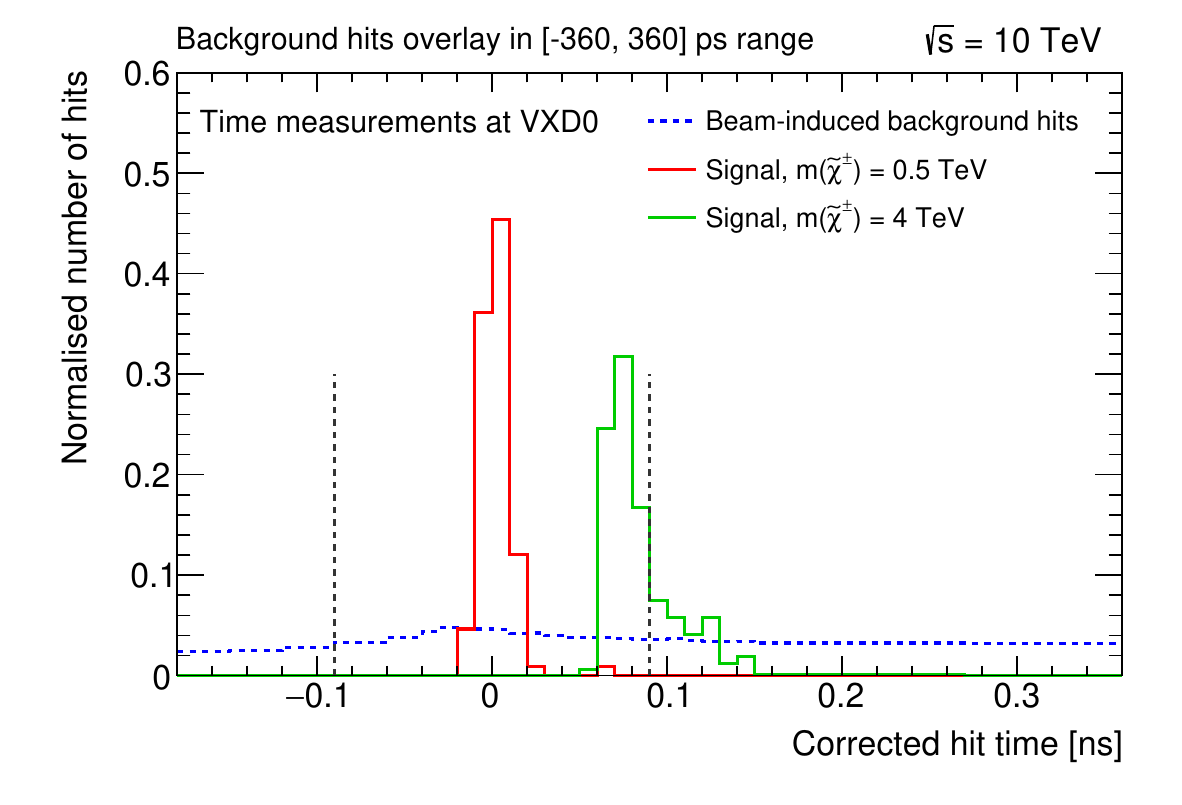}
		\caption{}
	\end{subfigure}%
    \begin{subfigure}[t]{0.495\textwidth}
    	\centering
    	\includegraphics[width=\textwidth]{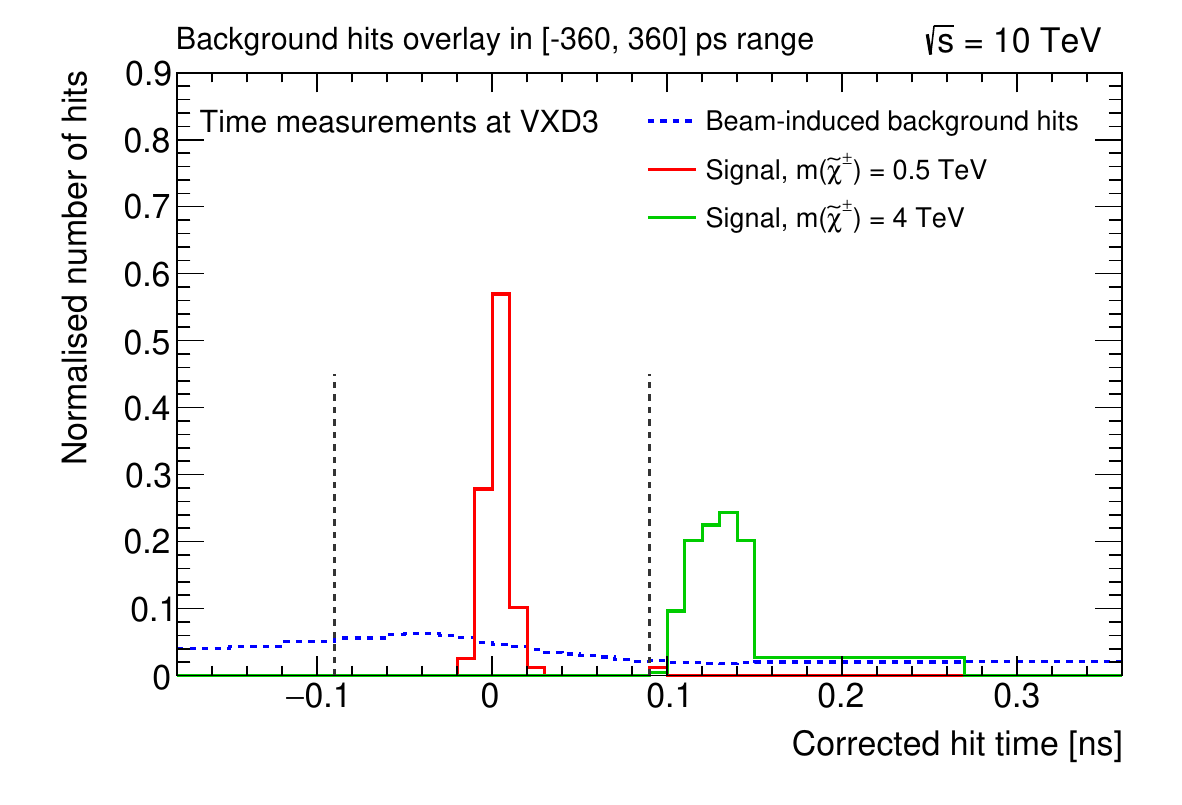}
		\caption{}
	\end{subfigure}
\end{center}
 \caption{Distribution of the measured detector hit times in the first layer, VXD0 (a), and the fourth layer, VXD3 (b), of the vertex detector corrected by the time of flight that a particle moving at the speed of light would need to reach the hit position if originating from the centre of the interaction region. The red and green lines represent different signal masses, for $\sqrt{s} = 10$~TeV collisions. The dashed blue line represents the distribution of the arrival time for hits from the BIB. The vertical dashed gray lines indicate the accepted time window for the nominal time selection cuts.}
 \label{fig:timing}
\end{figure}

The use of modern silicon sensors that provide both spatial and time information makes it possible to exploit the particle time-of-flight and arrival time on the sensors to reject those background components that are incompatible with the bunch crossing.  Figure~\ref{fig:timing} shows the hit arrival time, where the time measurement of each sensor is corrected by the time of flight that a particle moving at the speed of light would need to reach the detector, if originating from the centre of the interaction region. 

A symmetric window with a width of three times the time resolution of the silicon sensor, centered on the bunch crossing time, is applied to select the hits used for the track reconstruction. Future developments in tracking techniques exploiting the hit time information could partly remove the need for such selection cuts.
It is important to note that, while this selection is very powerful in rejecting out-of-time hits, it also implies a potentially significant loss of efficiency for high mass particles that move with relativistic velocities $\beta < 1$, especially for the detector layers at larger radii. Figure~\ref{fig:timing} shows how a \chipm with a mass of 4~TeV produced in $\sqrt{s}=10$~TeV collisions would never reach the fourth sensitive layer of the vertex detector in time for its energy deposits to be accepted by this time selection, causing a severe reconstruction inefficiency.
For this reason, a sequential procedure aimed at reconstructing tracks in the event considering only hits compatible with a certain interval of $\beta$, to recover the inefficiencies from the timing cuts, is assumed to be put in place. Such a procedure would allow to fully recover the signal track reconstruction efficiency while keeping a roughly constant level of BIB hits surviving the selection, at the cost of increased computational time due to the need to re-reconstruct the same event several times. Realistic limitations of the maximum duration of the read-out window of the tracking sensors, or the need to apply selections within the detector electronics could limit the detection efficiency for \chipm masses close to $\sqrt{s}/2$. For this reason, in the spirit of maximising the discovery opportunities at such a high-energy exploration machine, it is advisable to minimise as much as possible the use of timing requirements before the data is read out and saved to disk for offline analysis. In the following, it is assumed that the inefficiency due to the timing requirements can be minimised and it is therefore neglected. 

\begin{figure}[h]
\begin{center}
    \begin{subfigure}[t]{0.48\textwidth}
    	\centering
		\includegraphics[width=\textwidth]{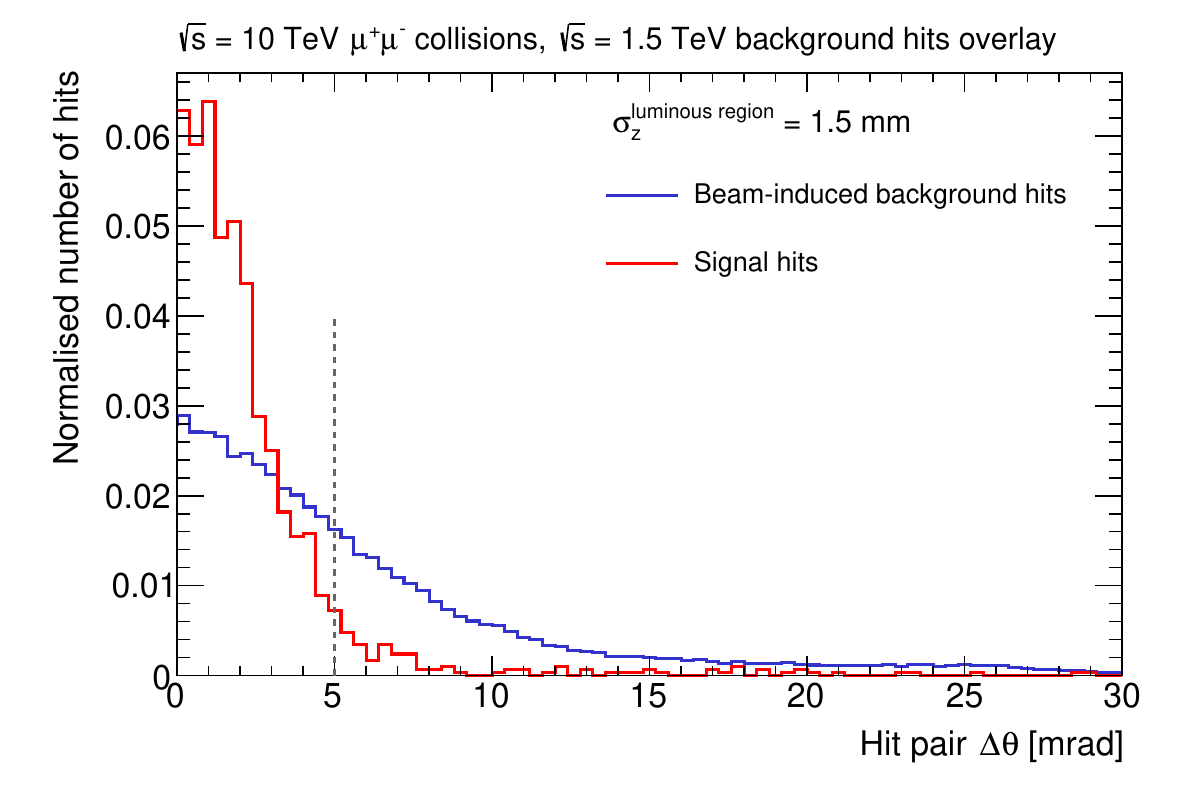}
		\caption{}
	\end{subfigure}%
    \begin{subfigure}[t]{0.48\textwidth}
    	\centering
    	\includegraphics[width=\textwidth]{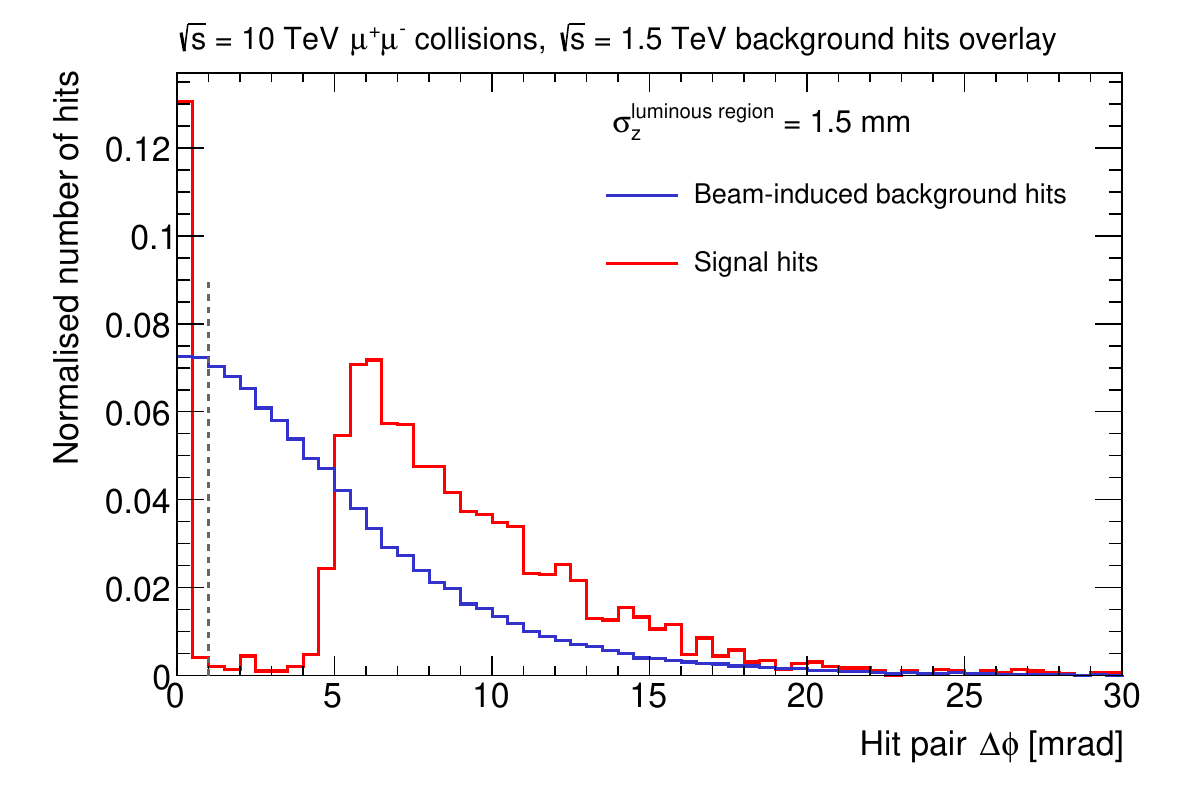}
		\caption{}
	\end{subfigure}
\end{center}
\caption{Distribution of the minimum polar (a) and azimuthal (b) angular difference between pairs of hits in the first double-layer of the vertex detector, in the barrel region. The dashed vertical lines indicate the maximum accepted value by the stub track selection. The double peak structure observed in (b) is due to the tracks corresponding to the low-momentum charged SM particle (typically a pion) which is produced in the \chipm decay and populates the higher $\Delta \phi$ region. Since reconstructing these low momentum tracks is not the target of our search, the corresponding detector hits are rejected by the chosen stub track selection.}
\label{fig:angle}
\end{figure}

The second handle to reject hits from the BIB is their spatial correlation in subsequent layers of the detector. The double-layer layout of the vertex detector can be exploited to reconstruct ``stub'' tracks from the pairs of hits in the neighbouring detector layers. The angular direction of such stub tracks can be exploited to reject pairs of hits that do not point back to the interaction region. The procedure is as follows: for each double-layer in the vertex detector, only the hits in the inner layer of the pair that have a corresponding hit in the outer layer within fixed thresholds in polar and azimuthal angle are retained. Assuming that particles propagate outward, for each of those retained inner hits, all hits within the same thresholds are retained as well.
Figure~\ref{fig:angle} illustrates the power of such a selection by showing the distribution of the polar angle difference in the innermost double-layer of the vertex detector for signal and BIB hits. Signal hit pairs in the outer double layers are characterised by smaller angular separations due to the longer distance from the interaction region. The hit pairs are accepted if they have a polar angular difference below $[10, 4, 2, 1]$~mrad (ordered outward) in the four double layers of the vertex detector and if they lie within 1~mrad in the azimuthal direction.
The use of pixel detector hit cluster shapes could be used for the same goal and could remove the need for double layers in at least the outermost part of the vertex detector, reducing the overall material budget. This is however not expected to affect the results of this work.

\begin{figure}[h]
 \centering
 \includegraphics[width=0.7\textwidth]{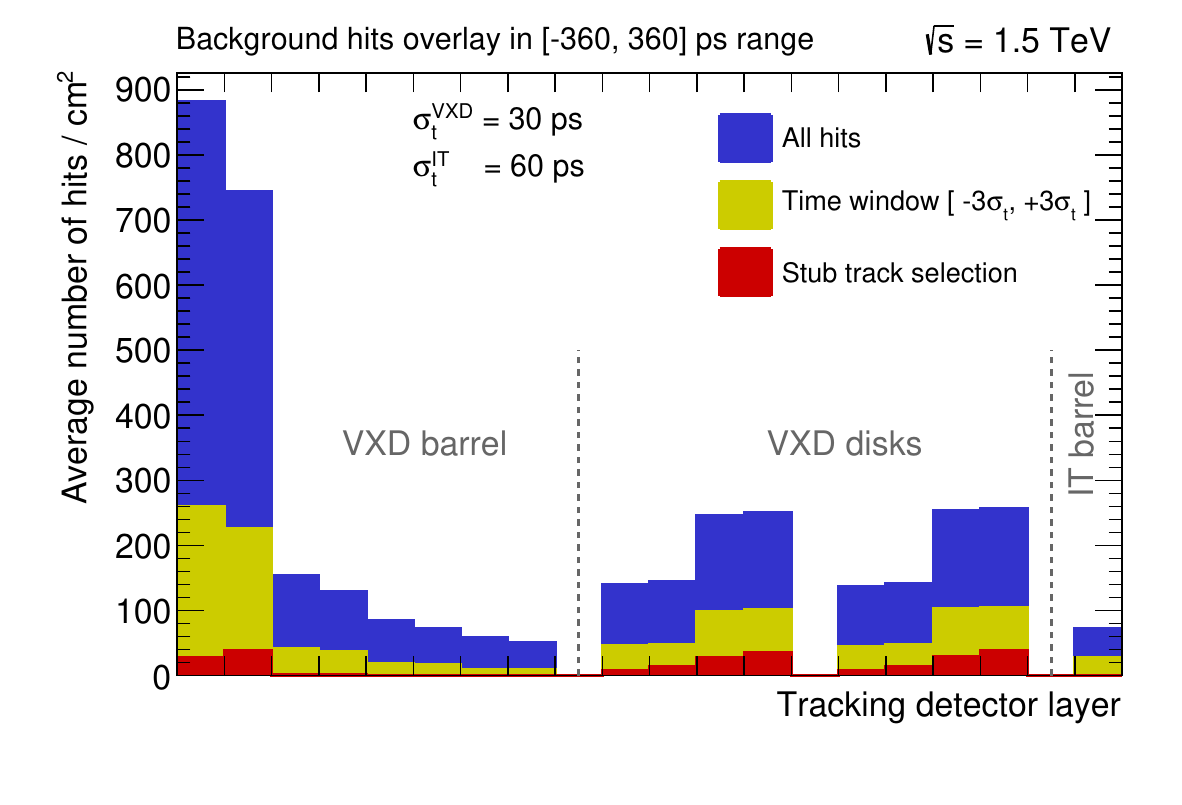}
 \caption{Hit density in the tracking detectors, shown separately for each layer of the sub-detectors considered in the tracklet reconstruction, for $\sqrt{s}=1.5$~TeV BIB. The empty bins highlight the boundaries between detector components, i.e. between the barrel and endcap disks sub-detectors and between the two sides of the detector in the disk sub-detector. The category ``all hits'' depends on the time window that is selected for the BIB overlay inside the simulation.}
 \label{fig:hitcutflow}
\end{figure}

The summary of the reduction in the tracker layer occupancy for the hit selections described in this section is shown in Figure~\ref{fig:hitcutflow}. The initial inclusive category has a rate that depends on the time window that is used for the BIB overlay inside the simulation: in this case, a symmetric window of 360~ps around the hard scatter interaction time. For the innermost layers, the combined selections reduce the hit density by up to two orders of magnitude to a maximum level of about 40 hits/cm$^2$, greatly simplifying the combinatorial problem behind track reconstruction.

\begin{figure}[h]
 \centering
 \includegraphics[width=0.7\textwidth]{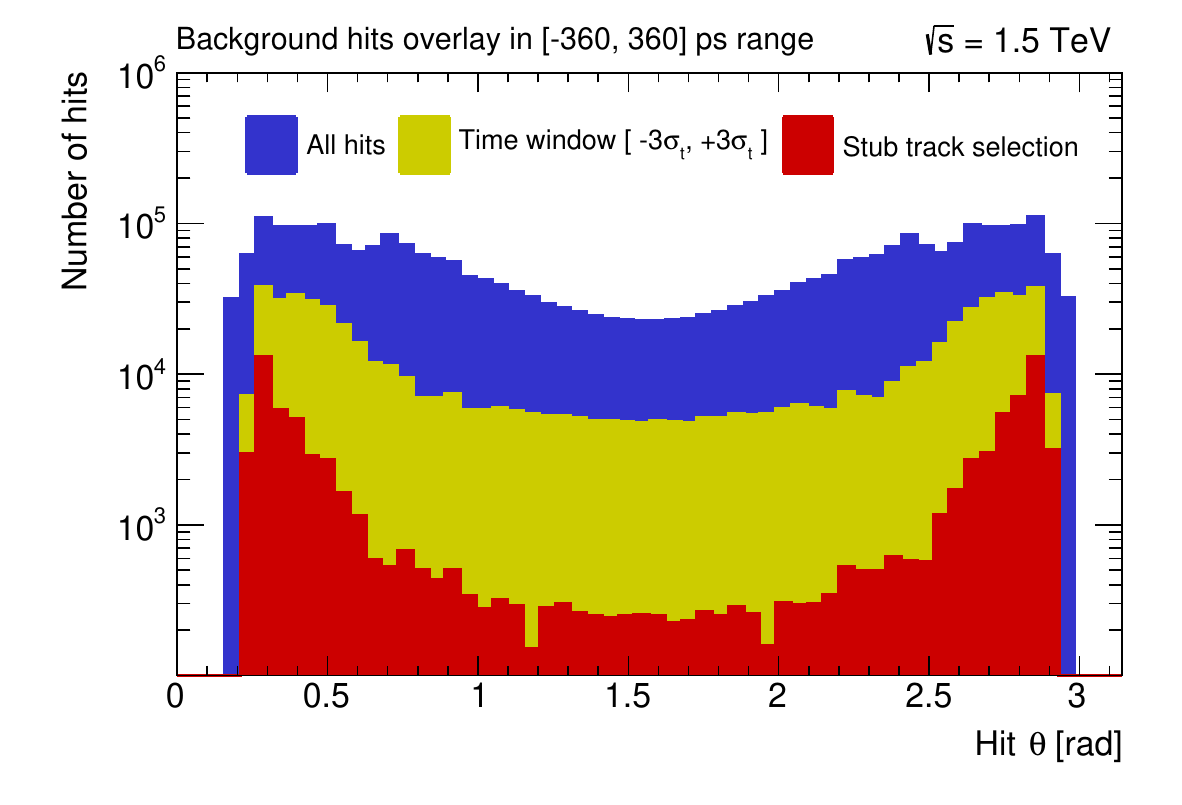}
 \caption{Hit density in the tracking detectors, shown as a function of the hit polar angle~$\theta$, for the BIB at $\sqrt{s}=1.5$~TeV. The category ``all hits'' depends on the time window that is selected for the BIB overlay inside the simulation.}
 \label{fig:hittheta}
\end{figure}

Figure~\ref{fig:hittheta} shows the number of hits in the tracking detectors as a function of the hit polar angle. After applying the full hit selection criteria, most surviving hits populate the regions of high or low $\theta$, which are expected to contain a relatively small number of signal events.

The combination of the two hit-level selections allows us to achieve an average BIB rejection factor of 10 in the subsystems of the tracking detectors used by this analysis, while retaining an average efficiency on signal hits of 83\%, strongly reducing the probability to reconstruct tracklets from random combinations of hits from the BIB.

\FloatBarrier
\subsection{Track-level rejection}
\label{subsec:track}

The hits satisfying the selections described in the previous section are used as input to the track reconstruction.
The resulting tracks are required to have at least four associated tracking detector hits (corresponding to the first two double-layers of the VXD) and are further required to satisfy a number of additional quality selections to reduce the number of fake tracks arising from the random alignment of detector hits.

\begin{figure}[h]
\begin{center}
    \begin{subfigure}[t]{0.495\textwidth}
    	\centering
		\includegraphics[width=\textwidth]{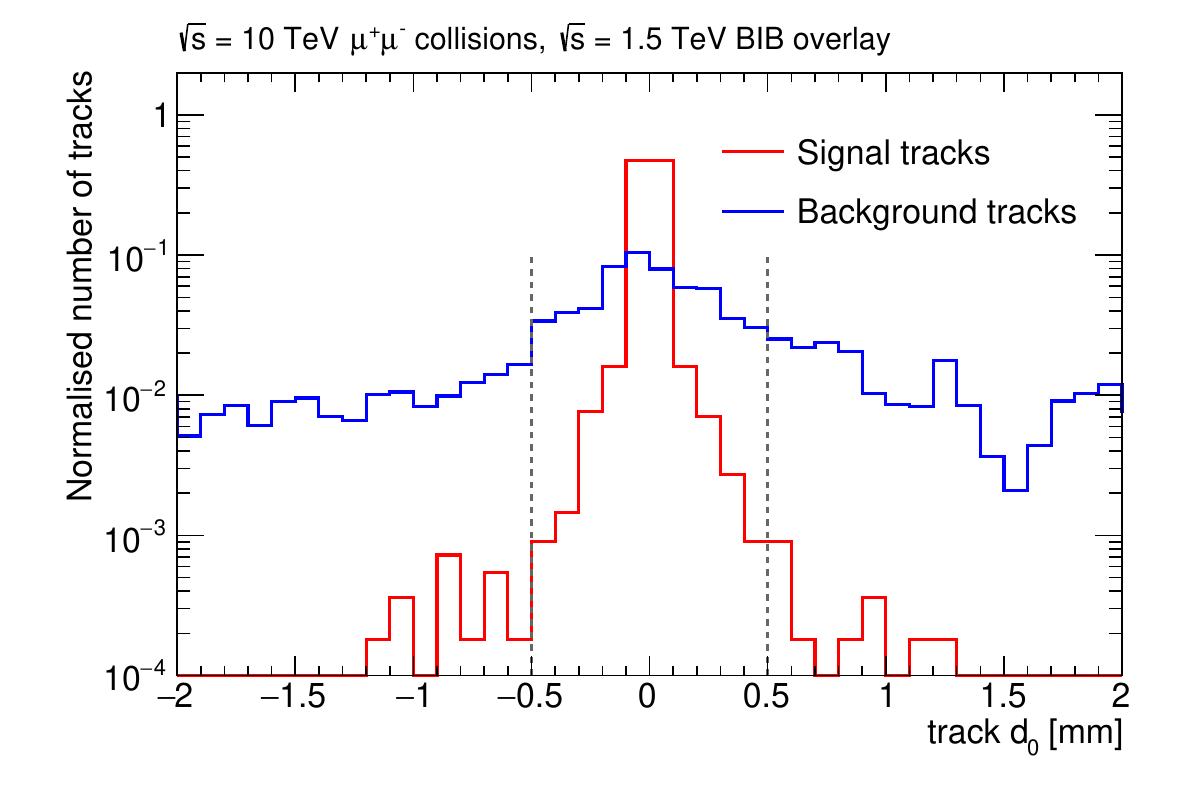}
		\caption{}
	\end{subfigure}
    %\begin{subfigure}[t]{0.48\textwidth}
    %	\centering
    %	\includegraphics[width=\textwidth]{figs/track_z0}
    %	\caption{\FM{fix x axis label}}
	%\end{subfigure}\\
    \begin{subfigure}[t]{0.495\textwidth}
    	\centering
		\includegraphics[width=\textwidth]{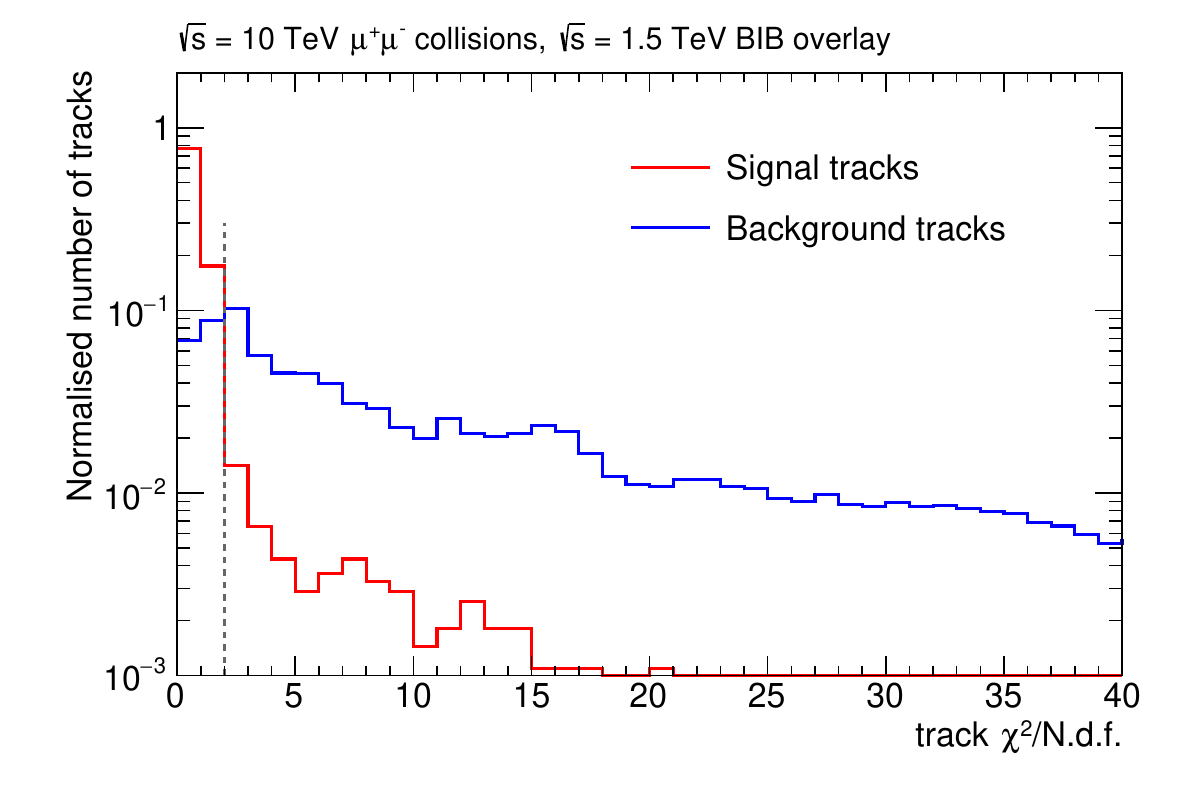}
		\caption{}
	\end{subfigure}\\
    \begin{subfigure}[t]{0.495\textwidth}
    	\centering
    	\includegraphics[width=\textwidth]{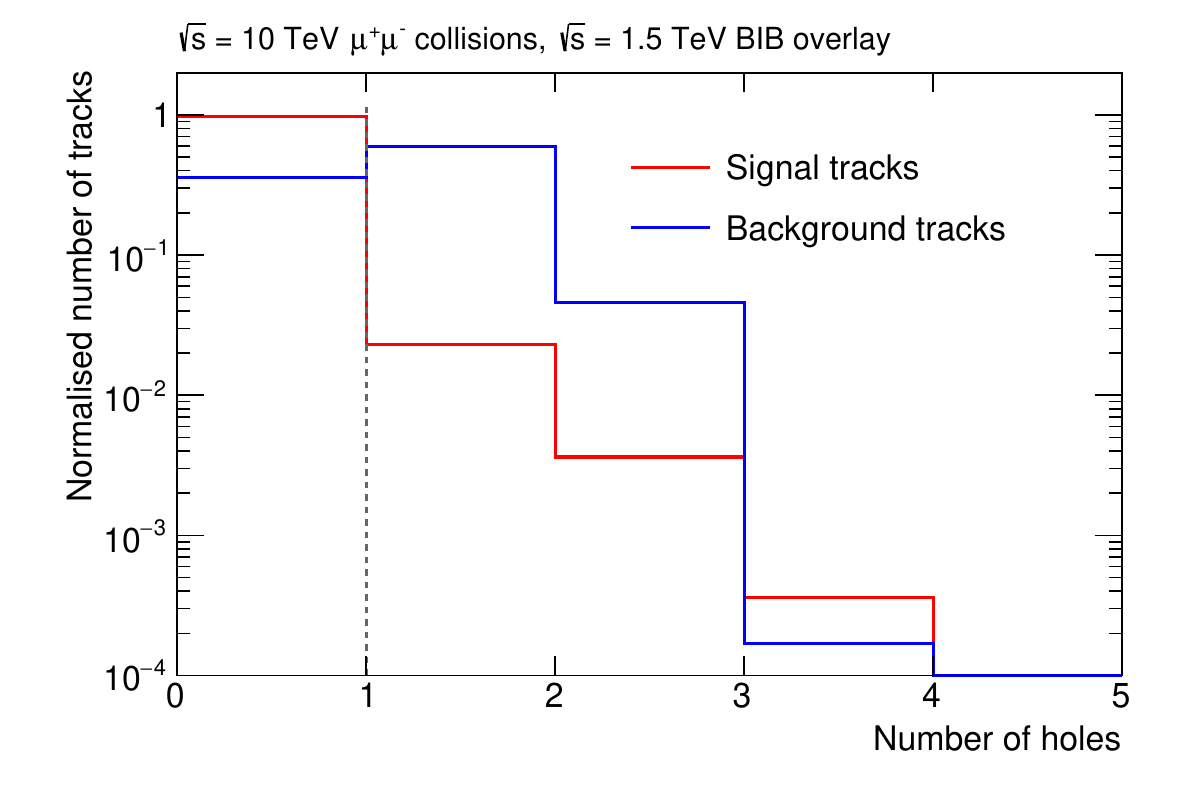}
    	\caption{}
	\end{subfigure}%
\end{center}
\caption{
Distributions of the track transverse impact parameter $d_0$ (a), the track fit quality $\chi^2/\text{N.d.f}$ (b), and number of holes (c). Signal tracks are presented in red, while background tracks originating from the BIB in blue. The histograms are normalised to unit area. The dashed grey lines represent the track selections applied to the track candidates. The signal track distributions are taken from a wino signal sample with $m(\chipm)=500$~GeV. A track is considered as signal if the hits matched to the generator-level \chipm compose more than 70\% of the total hits associated to the track. Background tracks are composed exclusively by hits from the BIB.
}
\label{fig:track_vars}
\end{figure}

The additional quality selections are shown in Figure~\ref{fig:track_vars} and proceed as follows. A selection is applied on the transverse ($d_{0}$) impact parameter relative to the centre of the interaction region to be smaller than 0.5~mm.
This selection is particularly powerful to reject fake tracks as these tend not to point back to the primary interaction region. Additional selection criteria on the longitudinal impact parameter could also be applied to further reduce the number of background tracks, but are not pursued in this study to avoid introducing a dependency on the expected interaction region spread in the longitudinal direction.
Tracks are further required to have a good quality of the fit by requiring the ratio between the $\chi_{\mathrm{max}}^2$ and the number of degrees of freedom (N.d.f.) to be below 5. A final selection requires the selected tracks to have no holes. Holes are defined as missing expected hits along the track direction. Spurious combinations of hits from uncorrelated particles are more likely to have missing hits than real particles that can only miss hits because of detector inefficiencies. 

Additional selections exploiting the timing information of the hits associated to a track were explored considering for example the average hit time corrected by the time of flight for the associated hits, or the largest corrected time difference between all hits associated to a track, finding possible additional discrimination power (up to a factor two of additional background rejection). However since these selections had the potential to introduce signal inefficiencies for low $\beta$ scenarios, it was decided not to pursue these as a part of the standard analysis.

\begin{figure}[h]
 \centering
 \includegraphics[width=0.7\textwidth]{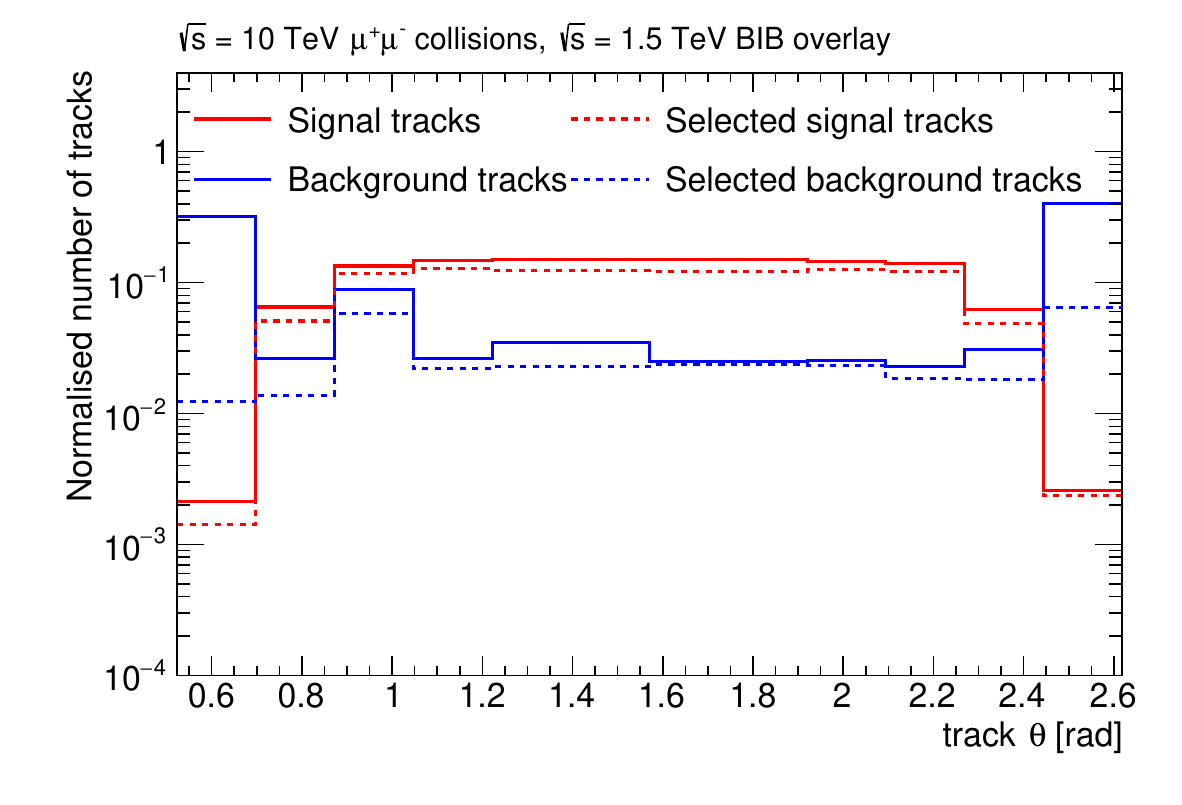}
 \caption{Distribution of the polar angle of the reconstructed tracks, and after all track quality requirements. Signal tracks are presented in red, while background tracks originating from the BIB in blue. The histograms are normalised to unit area. The dashed lines show the effect of the requirements imposed on the track $d_0$, $\chi^2/\text{N.d.f}$ and number of holes. The signal track distributions are taken from a wino signal sample with $m(\chipm)=500$~GeV. A track is considered as signal if the hits matched to the generator-level \chipm compose more than 70\% of the total hits associated to the track. Background tracks are composed exclusively by hits from the BIB.}
 \label{fig:tracklet_theta}
\end{figure}

Simulated events with the full BIB overlay were found to have a number of fake tracks satisfying all the above selections that is distributed according to a Poisson distribution with mean parameter 0.08. 
Figure~\ref{fig:tracklet_theta} shows the distribution of the polar angle for the reconstructed tracks from real \chipm and fakes for an inclusive selection as coming from the reconstruction software, and after all track quality requirements. Fake tracks are predominant in the forward region, where the hit multiplicity from the BIB dominates. The heavy \chipm instead tend to be produced centrally and the resulting track distribution reflects this behaviour.

The combination of these track-level selections achieves an average fake track rejection of about 5, while retaining an average signal track selection efficiency of 90\%. This allows us to attempt to select $\chipm\chimp$ events in the harsh muon collider environment.

\FloatBarrier
\subsection{Tracklet reconstruction efficiency}
\label{subsec:tracklets}

The track reconstruction efficiency for signal \chipm is estimated using fully simulated samples of $\chipm \chimp$ production with BIB overlay. Tracks are considered efficient if the hits matched to the generator-level \chipm compose more than 70\% of the total hits associated to the track and the track satisfies all quality selections defined in Section~\ref{subsec:track}. A track is considered ``reconstructable'' only if the \chipm traverses at least four detector layers. Excellent efficiency above 80\% is observed across the \chipm \pt spectrum for reconstructable tracks lying in the central region of the detector.

The final track-level selection is imposed as a ``disappearing'' condition. This requirement consists in vetoing tracks that have associated hits from a certain detector layer and beyond. For the study in this paper, tracks are vetoed if they have hits in the first layer of the IT (corresponding to a radius of 12.7 cm) or beyond. The tracks satisfying this final selection constitute the signal tracklets.

\begin{figure}[h]
 \centering
 \includegraphics[width=0.7\textwidth]{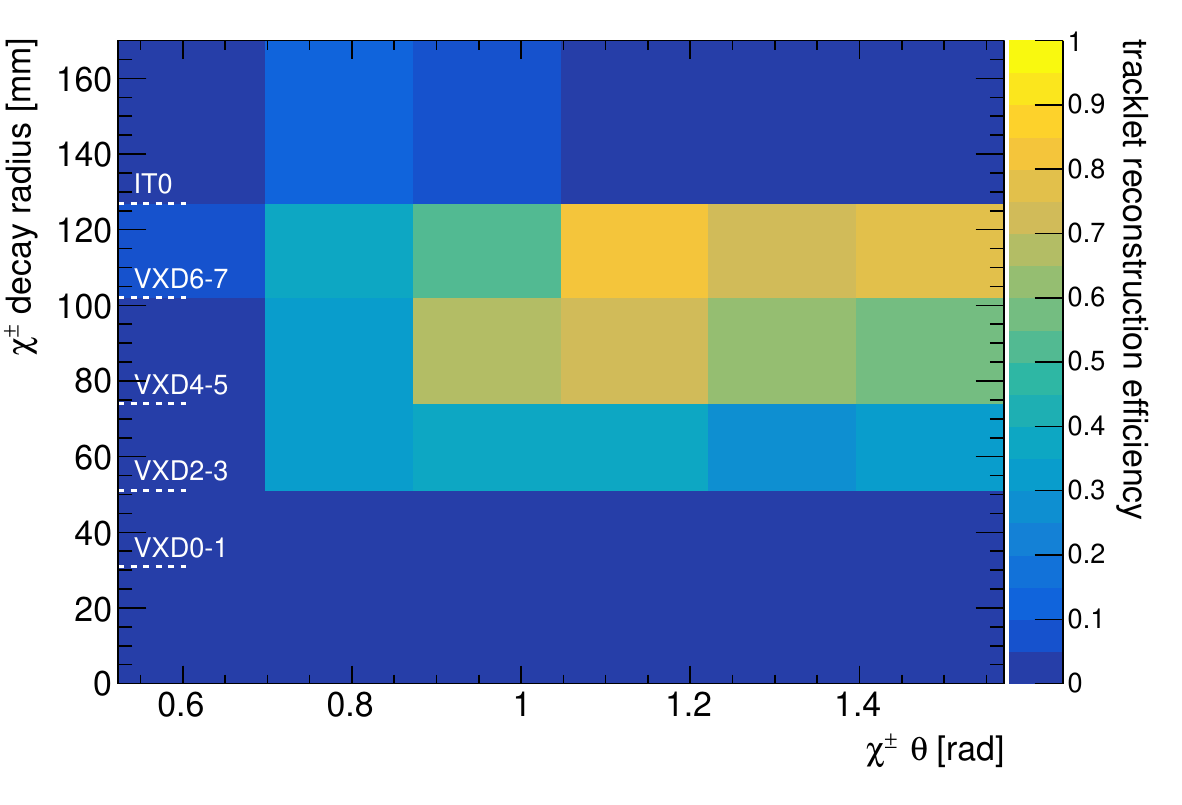}
 \caption{Tracklet reconstruction efficiency as a function of the generator-level \chipm polar angle $\theta$ and radial decay position. The decay radius bin boundaries correspond to the positions of the tracking detector layers in the barrel region, and are highlighted by the white dashed lines and text. The efficiency was found to be symmetric around $\pi/2$ within statistical uncertainties. The reconstruction efficiency decreases sharply at $\pi/6$ since hits at smaller polar angles are not considered in the reconstruction, as discussed in Section~\ref{sec:reco}.}
 \label{fig:eff_vs_theta_R}
\end{figure}

In order to extract a reconstruction efficiency parameterisation that is independent on the choices of \chipm mass and lifetime in the signal sample, a parameterisation is derived as a function of the generator-level \chipm polar angle $\theta$ and radial decay position. The resulting parameterisation is shown in  Figure~\ref{fig:eff_vs_theta_R} and used in the fast simulation of the signal events to reweight the event based on the probability of reconstructing the tracklets in the event.

These efficiencies are model independent and can be used to estimate the coverage of disappearing tracks at muon collider experiments using our detector layout for arbitrary models. Similarly, those readers interested in performing a more detailed study including a simplified simulation of the rates of BIB tracklets can build a simplified particle gun from the average track multiplicity per event, the polar angle distribution shown in  Figure~\ref{fig:tracklet_theta} and the \pT distribution presented in the following section.

\FloatBarrier

\section{Disappearing tracks at the muon collider}
\label{sec:analysis}
In this section we discuss our analysis strategy, having established that the BIB can be safely reduced to manageable levels. To begin with, we describe in Section~\ref{subsec:selection} our event selection, which is based on establishing signal regions (SRs) that depend on the number of reconstructed tracklets and on possible additional requirements on the presence of additional photon radiation. Our results showing the MuC sensitivity curves, overlayed with expectation from other future colliders and the HL-LHC reach, are presented in Section~\ref{subsec:results}.

\subsection{Event selection}
\label{subsec:selection}

In order to be considered for this analysis, events must contain at least a reconstructed tracklet and no reconstructed leptons or jets.
Two SRs were optimised to maximise the discovery potential. The first selection, labelled as \SRotp, relies on the identification of a single energetic tracklet. The second selection, labelled as \SRttp, is instead aimed at higher signal purity and requires the identification of a pair of energetic tracklets in the event.

The Monte Carlo simulated event samples described in Section~\ref{sec:sim} are used to predict the backgrounds in the SRs. There are two main background contributions: SM particles that are reconstructed as tracklets, and events that contain fake tracklets. The SM particles reconstructed as tracklets are typically hadrons scattering in the detector material or electrons undergoing bremsstrahlung. In the following, the contribution from the former is assumed to be negligible. Recent LHC searches~\cite{Sirunyan:2020pjd} have demonstrated that these backgrounds can be suppressed to a negligible level exploiting calorimeter energy vetoes, with no sizeable signal efficiency loss.
The contribution that arises from events that contain fake tracklets was modelled in both signal and background samples overlaying fake tracklets extracted from simulated events to the hard scatter events in the event reconstruction step with a multiplicity following a Poisson distribution with the mean parameter extracted from the simulation. 

\begin{figure}[h]
\begin{center}
    \begin{subfigure}[t]{0.495\textwidth}
    	\centering
		\includegraphics[width=\textwidth]{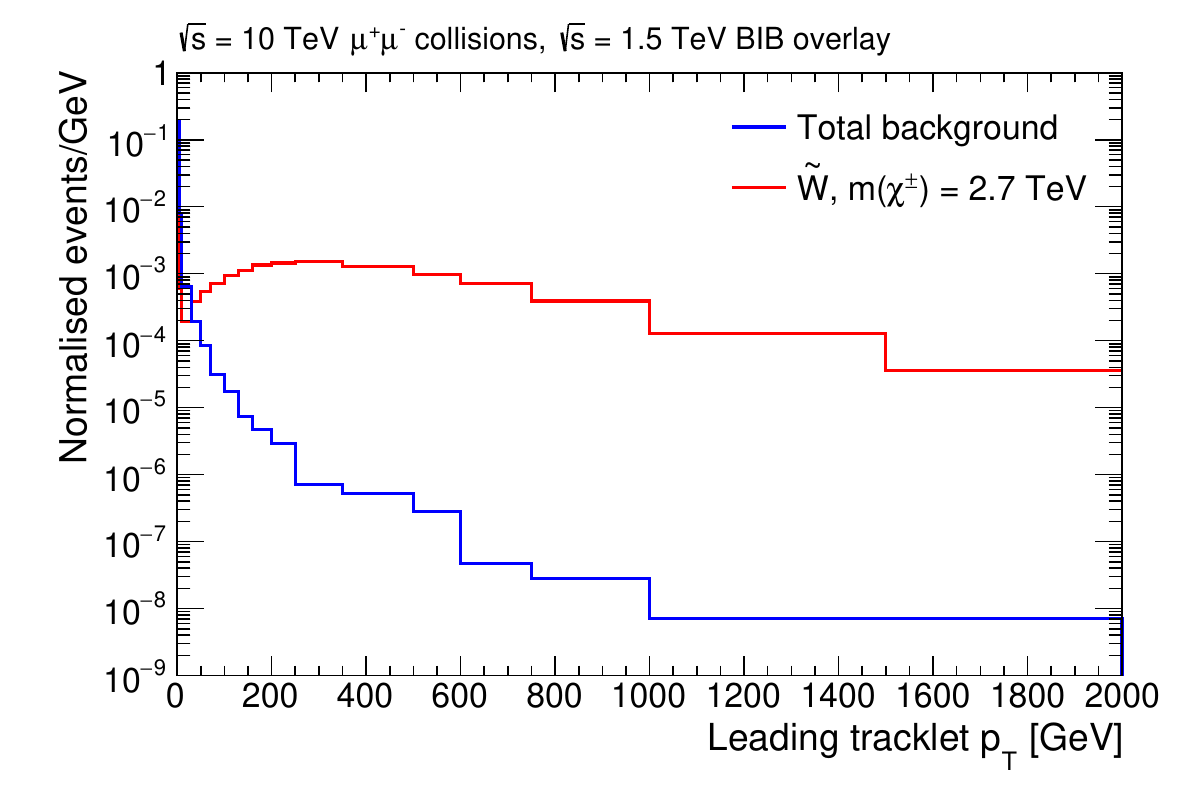}
		\caption{}
	\end{subfigure}
    \begin{subfigure}[t]{0.495\textwidth}
    	\centering
    	\includegraphics[width=\textwidth]{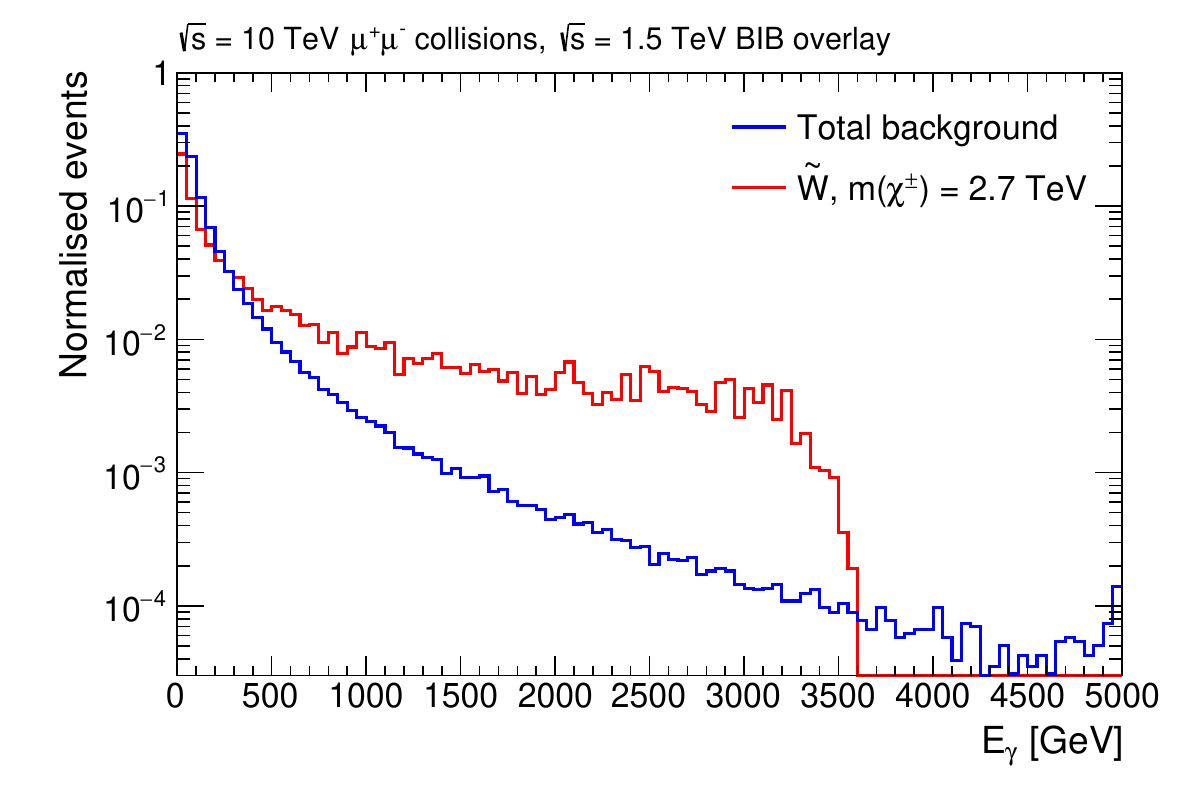}
    	\caption{}
	\end{subfigure}%
\end{center}
\caption{Distributions of the leading tracklet transverse momentum (a) and the leading photon energy (b) in $\sqrt{s}=10$~TeV muon collision events with at least a tracklet satisfying the requirements described in Section~\ref{sec:bib}.}
\label{fig:kinematics}
\end{figure}

In most events the \chipm are produced back to back and yield little momentum imbalance. The key discriminant variable in \SRotp for the rejection of the backgrounds is the transverse momentum of the reconstructed tracklets. Figure~\ref{fig:kinematics} shows the distributions of the reconstructed leading tracklet transverse momentum and the leading photon energy in events with at least a disappearing track satisfying the requirements described in Section~\ref{sec:bib} in $\sqrt{s}=10$~TeV muon collisions. While the disappearing track transverse momentum is a strongly discriminant variable, the leading photon energy was found to have relatively poor discrimination power and the minimum reconstructable photon energy threshold was chosen for the event selection in the conservative scenario.

\begin{figure}[h]
\centering
\includegraphics[width=0.495\textwidth]{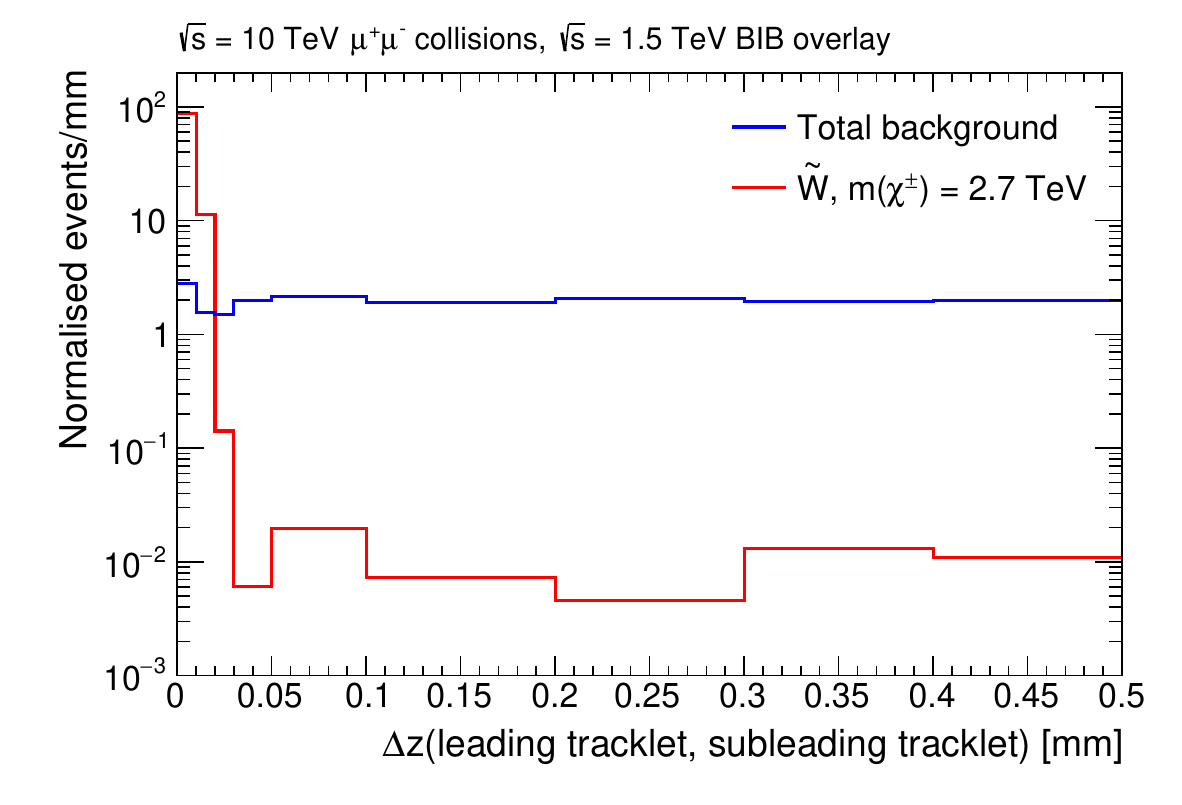}
\caption{Distributions of the $\Delta z$ between tracklets in $\sqrt{s}=10$~TeV muon collision events with at least two tracklets satisfying the requirements described in Section~\ref{sec:bib}.}
\label{fig:kinematics_deltaZ}
\end{figure}

In the case of \SRttp the distance between the two tracklets along the beam axis $\Delta z$, shown in Figure~\ref{fig:kinematics_deltaZ}, provides additional rejection power against fake tracklets and allows us to relax the requirements on the tracklet \pT in these selections. In order to maximise the signal acceptance of this selection, one of the two tracklets is required to satisfy the disappearing condition as described in Section~\ref{subsec:track}, while the second tracklet is allowed to have a longer decay length and is required not to have hits beyond the middle layer of the outer tracker detector (115.3~cm). The reconstruction efficiency for such longer tracklets has been conservatively assumed to be equal to that observed at a radius of 12~cm from Figure~\ref{fig:eff_vs_theta_R}.
In both \SRotp and \SRttp, the leading tracklet is required to lay within $\theta~[2/9\pi, 7/9\pi]$ in order to reject the fake tracklets that mostly populate the forward regions along the beam axis. Table~\ref{tab:SRdef} summarises the full SR selections used in the analysis at $\sqrt{s}=10$~TeV. 

An alternative scenario was also considered during the optimisation, assuming that data-taking at the muon collider could proceed without need for online event selection and that the full detector information could be read out for each bunch crossing. This scenario would allow to remove the requirement on the presence of a photon from initial or final state radiation, and the related signal acceptance loss. However, in this scenario the background yields depend on the fraction of time with the machine operating for physics data-taking. This dependence is due to the fact that the energy deposits due to the BIB will be recorded by the detector for each bunch-crossing in the machine, even in the cases when no hard scatter occurs in the colliding muon bunches. In order to produce a background estimate, we assumed a fraction of time in physics data-taking of 40\% across the scheduled five-year runs, based on typical values for future colliders~\cite{Bordry:2018gri}. Similarly, a data-taking efficiency of 90\% was assumed during the data-taking time. Under these assumptions, it was found that relatively simple selections as those used in \SRotp and \SRttp would be ineffective.

\begin{table}[!htb]
\def\arraystretch{1.2}
\begin{center}
\begin{tabular}{l r r}
\noalign{\smallskip}\hline\noalign{\smallskip}
Requirement / Region                            & \SRotp      & \SRttp \\
\noalign{\smallskip}\hline\noalign{\smallskip}
Vetoes       & \multicolumn{2}{c}{\textrm leptons and jets}  \\
Leading tracklet \pt~[GeV]   &  $>300$ & $>20$  \\
Leading tracklet $\theta$~[rad] &  \multicolumn{2}{c}{$[2/9\pi, 7/9\pi]$}  \\
Subleading tracklet \pt~[GeV]  &  - & $>10$ \\
Tracklet pair $\Delta z$ [mm]  &  - & $<0.1$ \\
Photon energy [GeV]  &  $>25$ & $>25$ \\
\hline
\end{tabular}
\end{center}
\caption{\label{tab:SRdef} Definition of the signal regions used in the analysis.}
\end{table}

\FloatBarrier
\subsection{Results}
\label{subsec:results}

The expected SR yields, for the total background contribution and for the wino and higgsino thermal targets, are reported in Table~\ref{tab:SRyields} for an integrated luminosity of 10~ab$^{-1}$ of \mbox{$\sqrt{s}=10$}~TeV muon collisions. 
%The expected number of events in the wino and higgsino scenarios are similar due to the disappearing condition, which causes a significant inefficiency for signals with long lifetimes.

\begin{table}[h]
\def\arraystretch{1.2}
\begin{center}
\begin{tabular}{l r r }
\noalign{\smallskip}\hline\noalign{\smallskip}
     & \SRotp      & \SRttp \\
\noalign{\smallskip}\hline\noalign{\smallskip}
Total background       & $187.8 \pm 0.6$ & $0.16 \pm 0.05 $  \\
$\tilde{W}$, 2.7 TeV, $\tau = 0.2$~ns  & $313 \pm 5$ & $168 \pm 2$ \\
$\tilde{H}$, 1.1 TeV, $\tau = 0.02$~ns & $53.0 \pm 0.7$ & $3.92 \pm 0.05$ \\
\hline
\end{tabular}
\end{center}
\caption{Expected numbers of events in the signal regions for an integrated luminosity of 10~ab$^{-1}$ of $\sqrt{s}=10$~TeV muon collisions. Statistical uncertainties due to the number of generated events are given. \label{tab:SRyields}}
\end{table}

\begin{figure}[h]
\begin{center}
    \begin{subfigure}[t]{0.48\textwidth}
    	\centering
		\includegraphics[width=\textwidth]{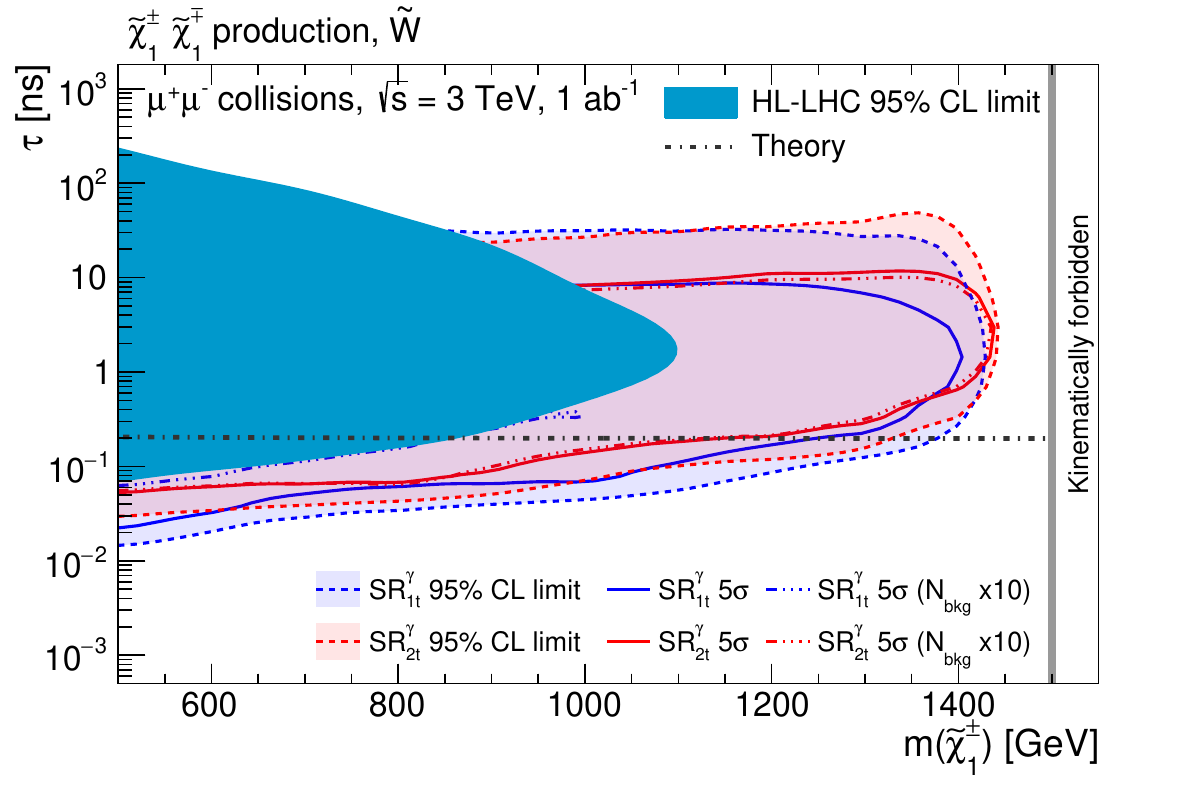}
		\caption{}
	\end{subfigure}
    \begin{subfigure}[t]{0.48\textwidth}
    	\centering
		\includegraphics[width=\textwidth]{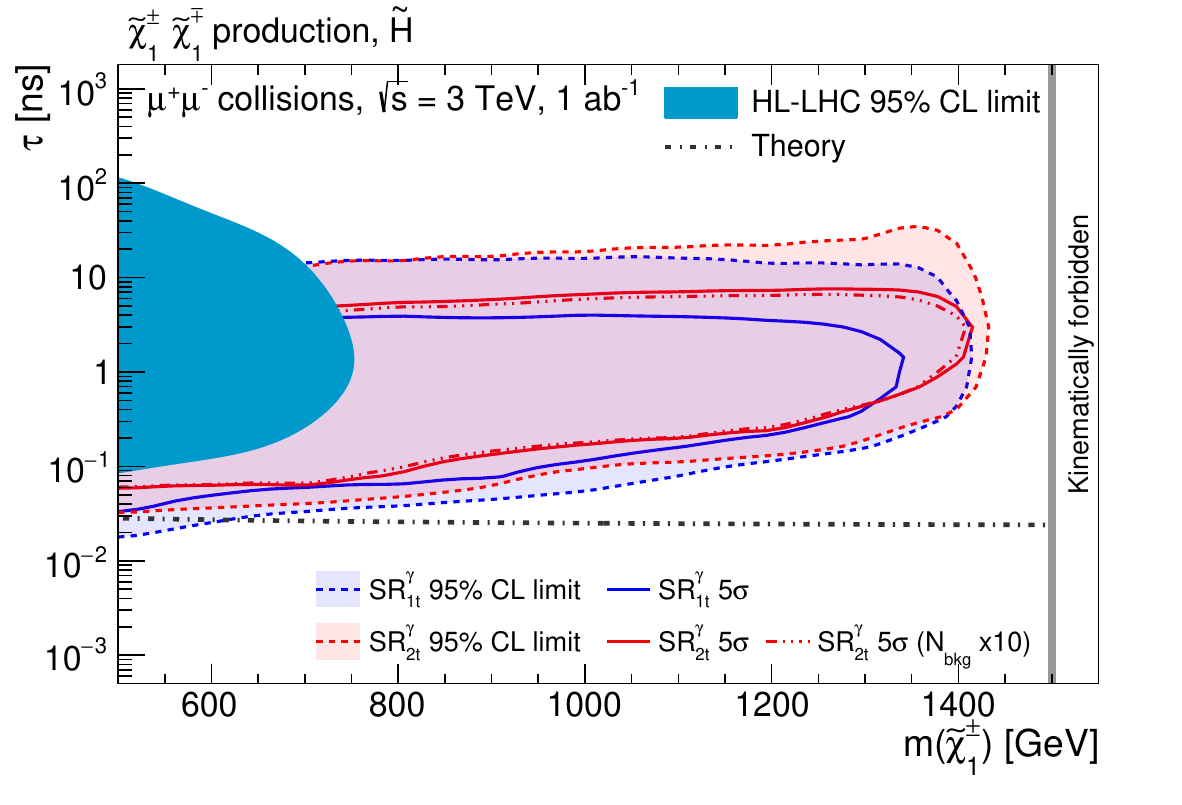}
		\caption{}
	\end{subfigure}\\ \smallskip
    \begin{subfigure}[t]{0.48\textwidth}
    	\centering
		\includegraphics[width=\textwidth]{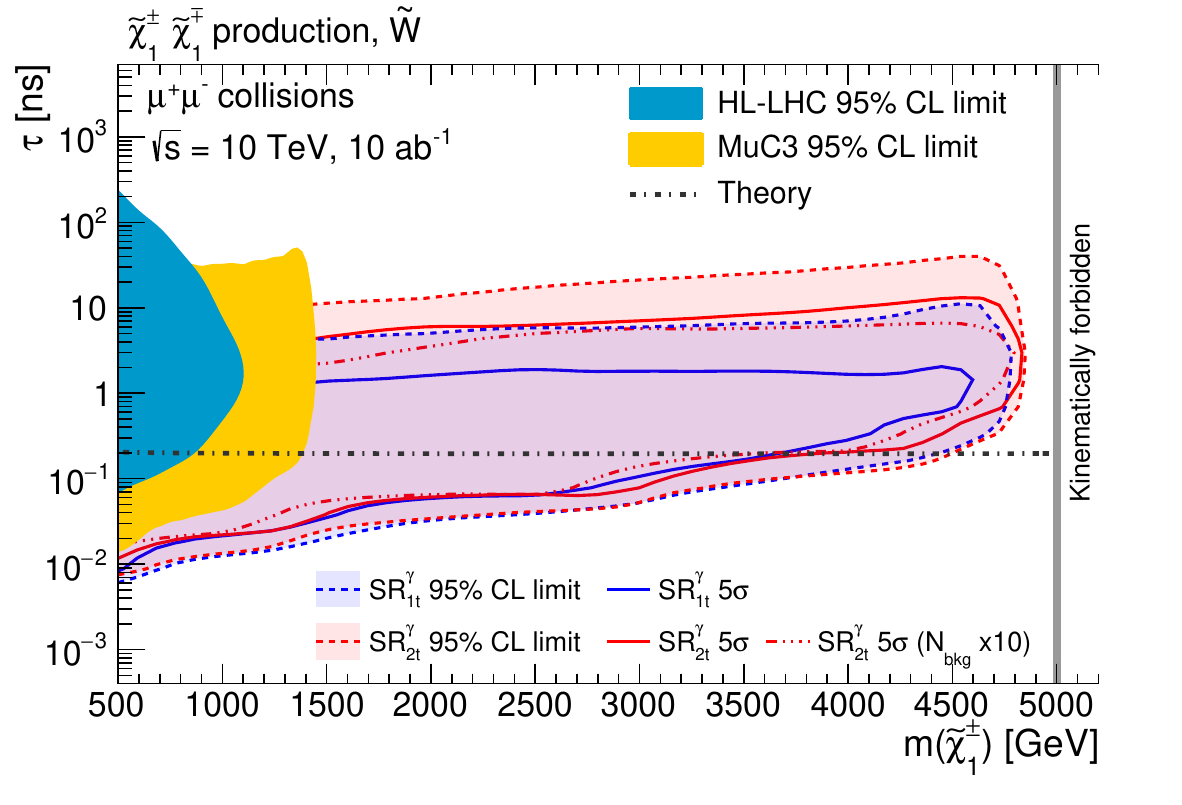}
		\caption{}
	\end{subfigure}
    \begin{subfigure}[t]{0.48\textwidth}
    	\centering
		\includegraphics[width=\textwidth]{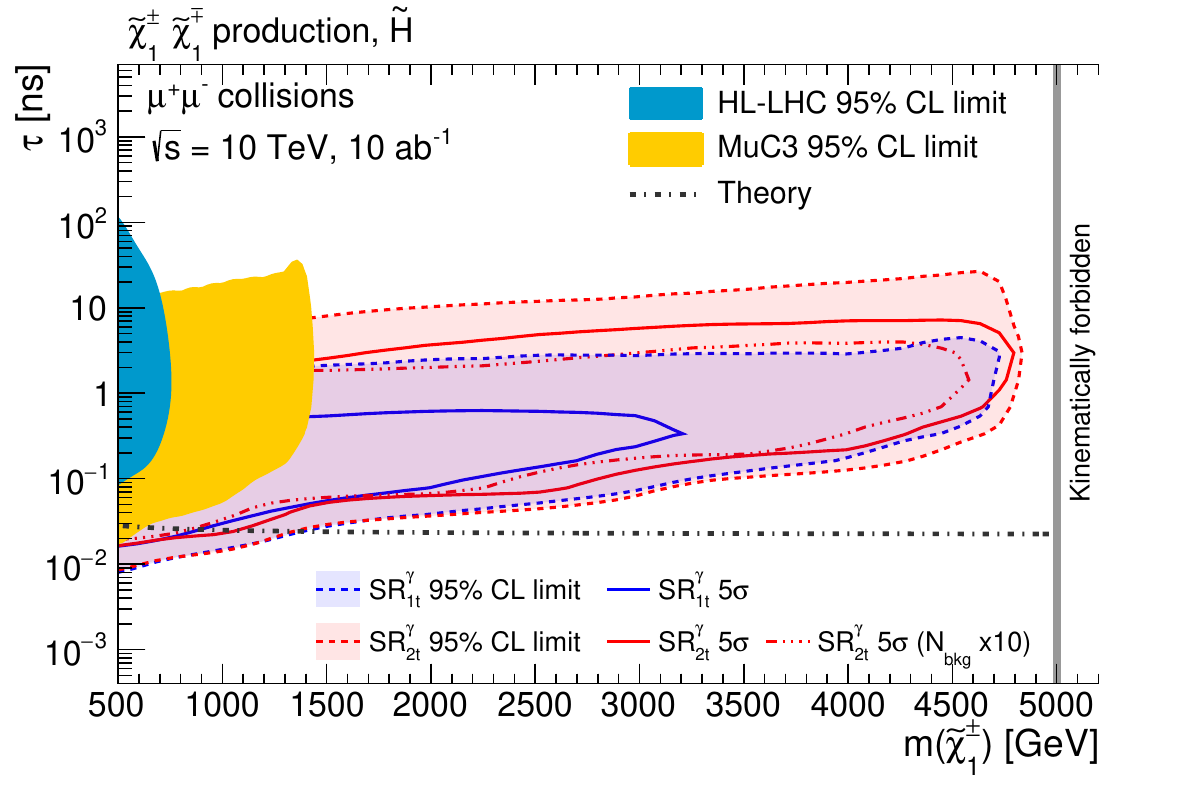}
		\caption{}
	\end{subfigure}
\end{center}
 \caption{Expected sensitivity using 1 ab$^{-1}$ of 3 TeV or 10 ab$^{-1}$ of 10 TeV $\mu^{+}\mu^{-}$ collision data as a function of the \chipm mass and lifetime. Models including $\chipm\chimp$ are considered assuming pure-wino scenarios (a and c) and pure-higgsino scenarios (b and d). The \chipm lifetime as a function of the \chipm mass is shown by the dashed grey line: in the pure-wino scenario it was calculated at the two-loops level \cite{Ibe:2012sx}, in the pure-higgsino scenario it was calculated at the one-loop level \cite{Thomas:1998wy,Cirelli:2005uq}.}
 \label{fig:sensitivity_m_vs_tau}
\end{figure}

Sensitivity curves are shown in  Figure~\ref{fig:sensitivity_m_vs_tau} as a function of the \chipm mass and lifetime.
A set of likelihoods is built for each signal mass and lifetime hypothesis and SR selection.
Each likelihood is a product of a Poisson probability density function, describing the observed number of events in the SR, and a single Gaussian probability density function distribution that describes a nuisance parameter associated with the total background systematic uncertainty. A systematic uncertainty of 30\% (100\%) on the total background prediction has been assumed for \SRotp (\SRttp) for the $\sqrt{s}=3$~TeV data-taking run. When considering the $\sqrt{s}=10$~TeV data-taking run, the systematic uncertainty on the total background prediction in \SRotp has been reduced to 10\%.
The pyhf software package~\cite{pyhf,pyhf_joss} was used to evaluate the discovery significance from the expected discovery $p$-value and to set limits at 95\% CL using the CLs method~\cite{Read:2002hq}. Additional lines show the sensitivity of the conservative scenario inflating the background estimates by an order of magnitude. Note that in \SRttp\ the expected background is relatively small. Hence enlarging the background in that region by an order of magnitude does not have a significant impact on the sensitivity. A proper treatment of the small background is achieved due to the use of Poissonian statistics.
The sensitivity is shown separately for the $\sqrt{s}=3$~TeV and $\sqrt{s}=10$~TeV data-taking runs, and for wino and higgsino multiplets. Available HL-LHC prospects \cite{ATL-PHYS-PUB-2018-031, Strategy:2019vxc} are also included for comparison. Limits at 95\% CL extracted from the $\sqrt{s}=3$~TeV data-taking are overlaid on the $\sqrt{s}=10$~TeV discovery prospects.

In the most favourable scenarios, the analysis of the full muon collider data set is expected to allow the discovery \chipm masses up to value close to the kinematic limit of $\sqrt{s}/2$. The interval of lifetimes covered by the experimental search directly depends on the layout of the tracking detector and the choices made in the reconstruction and identification of the tracklets. In particular, the setup presented in this work should allow to probe lifetimes down to $10^{-2}$~ns for a wide range of \chipm masses. Given the requirement of having at least four hits associated to the tracklet, this sensitivity limit depends of the radial position of the fourth tracking layer. In the detector used for this study, this corresponds to the position of the second double layer of the vertex detector.  The drop in sensitivity in Figure~\ref{fig:sensitivity_m_vs_tau} for shorter lifetimes as the chargino mass increases is caused by the reduced Lorentz boost of the
produced \chipm which decreases the probability for them to reach the second double layer and satisfy the minimum requirements. Most notably this affects the reach for pure higgsino models, as the shorter expected lifetime falls into this region with degraded sensitivity. In pure wino models, it is expected to be able to discover \chipm with lifetimes up to about 10~ns, in a large range of masses up to 4.5~TeV. The sensitivity at long lifetimes depends on the radial position of the tracking detector used to implement the disappearing condition and could be trivially extended by considering longer tracklets and imposing the disappearing condition only in the last layers of the tracking detector.
It is worth noting that even the most pessimistic scenarios considered are expected to potentially discover the thermal wino scenario at a 10~TeV MuC.

\begin{figure}[h]
 \centering
 \includegraphics[width=0.7\textwidth]{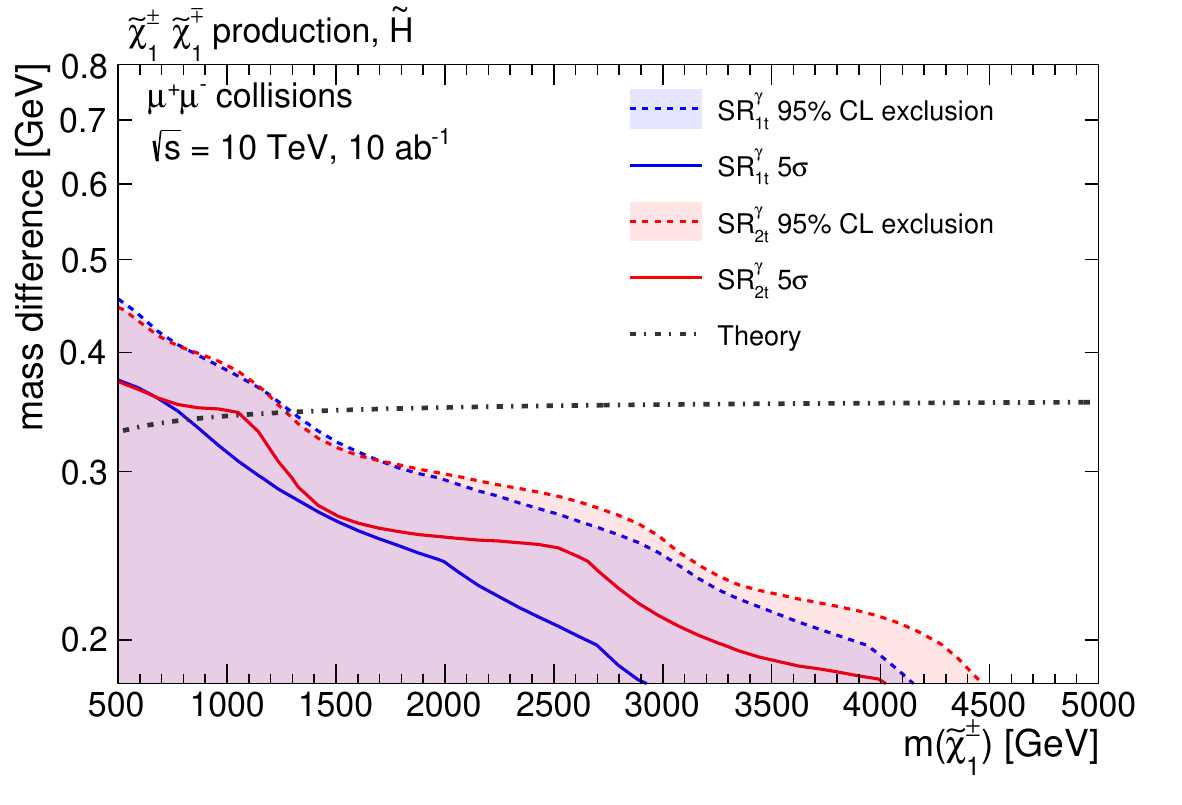}
 \caption{Expected sensitivity using 10 ab$^{-1}$ of 10 TeV $\mu^{+}\mu^{-}$ collision data as a function of the \chipm mass and mass difference with the lightest neutral state, assuming a pure-higgsino scenario. The mass splitting as a function of the \chipm mass is shown by the dashed grey line and was calculated at the one-loop level \cite{Thomas:1998wy,Cirelli:2005uq}.
 }
 \label{fig:sensitivity_deltam_hino}
\end{figure}

For the higgsino models, we provide also the expected sensitivity as a function of the mass of the \chipm and its mass splitting with respect to the lightest neutral state. The results are shown in Figure~\ref{fig:sensitivity_deltam_hino}.

\begin{figure}[h]
 \centering
 \includegraphics[width=0.7\textwidth]{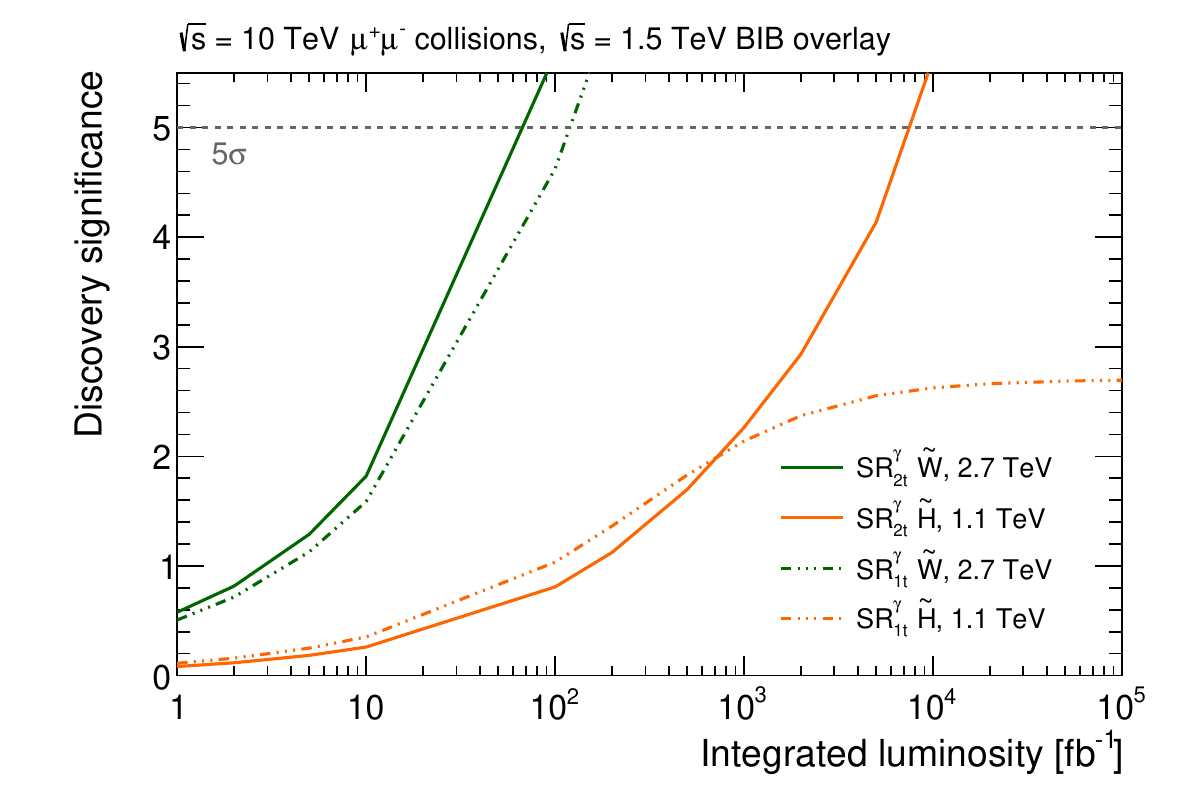}
 \caption{Expected discovery significance as a function of the integrated luminosity in 10~TeV $\mu^{+}\mu^{-}$ collisions, shown separately for two benchmark signals and each SR selection.
 }
 \label{fig:sensitivity_vs_lumi}
\end{figure}

Furthermore, we present in Figure~\ref{fig:sensitivity_vs_lumi} the expected discovery significance, in the $\sqrt{s}=10$~TeV configuration, as a function of the integrated luminosity for the wino and higgsino thermal targets. Under the nominal background hypothesis, we expect to be able to discover the wino thermal target with a minimum integrated luminosity of about 70~fb$^{-1}$ at $\sqrt{s}=10$~TeV, or approximately $1/100$ of the target selected for this study. The higgsino thermal target could be discovered with an integrated luminosity of approximately 8~ab$^{-1}$.

\begin{figure}[h]
 \centering
 \includegraphics[width=1.0\textwidth]{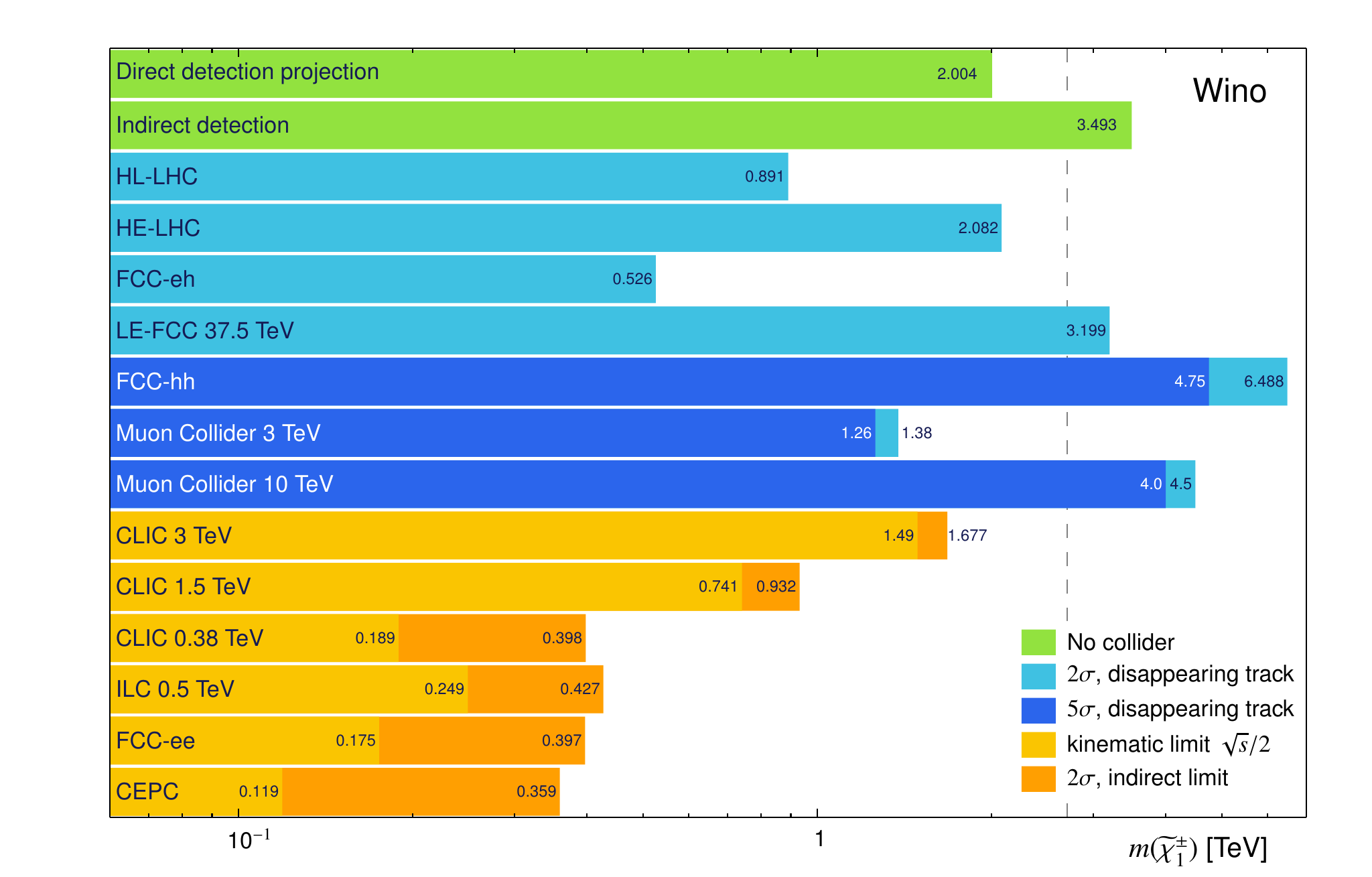}
 \caption{Summary of the sensitivity to pure wino models at future experimental facilities. The results for other facilities are taken from Refs.~\cite{Strategy:2019vxc,Saito:2019rtg}.
 }
 \label{fig:sensitivity_mass_other_wino}
\end{figure}

\begin{figure}[h]
 \centering
 \includegraphics[width=1.0\textwidth]{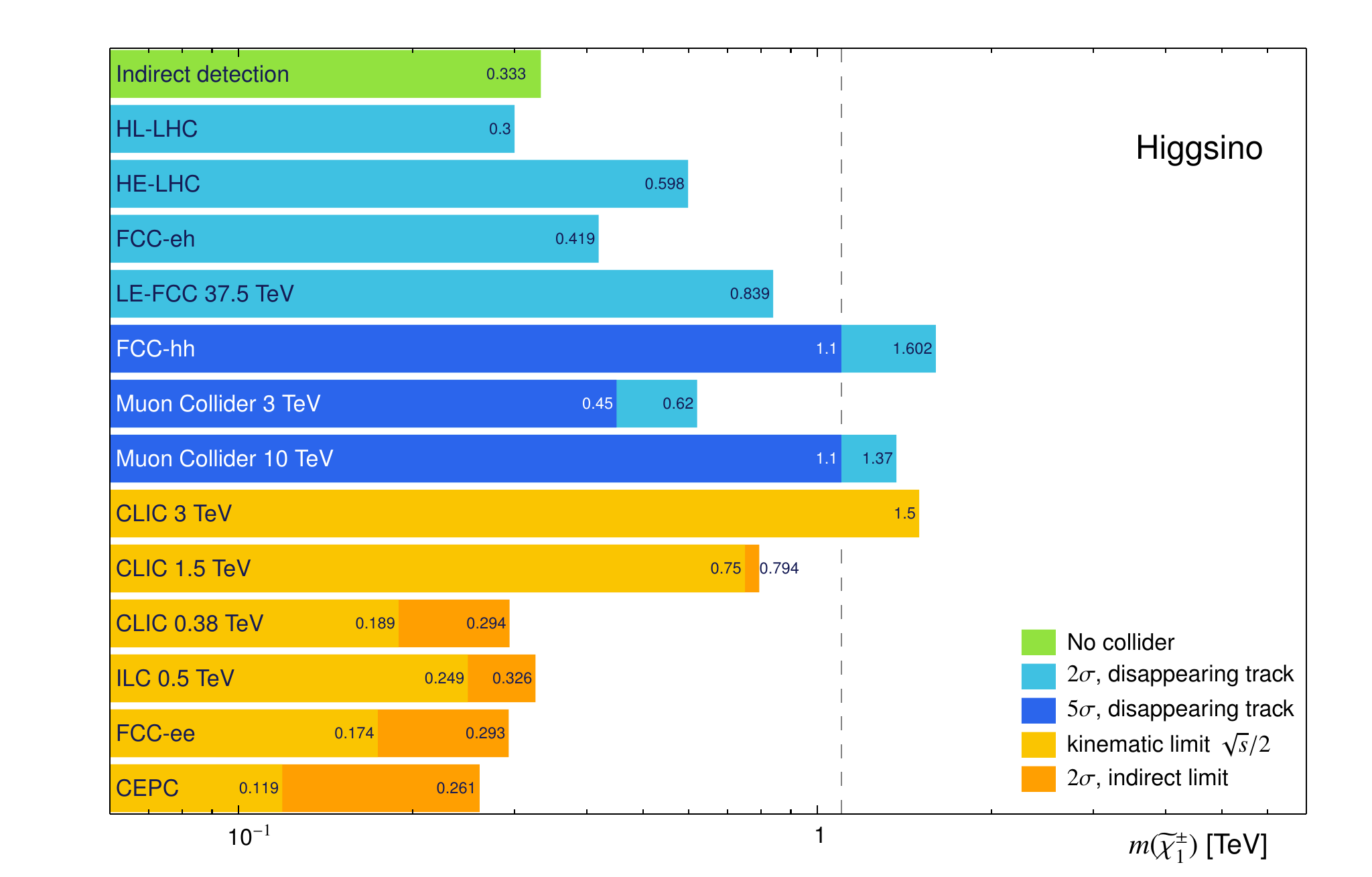}
 \caption{Summary of the sensitivity to pure higgsino models at future experimental facilities. The results for other facilities are taken from Refs.~\cite{Strategy:2019vxc,Saito:2019rtg}.}
 \label{fig:sensitivity_mass_other_hino}
\end{figure}

Finally, we present in Figure~\ref{fig:sensitivity_mass_other_wino} and~\ref{fig:sensitivity_mass_other_hino} a comparison of the MuC sensitivity to several other future proposed facilities. As these results come from different sources, we present exclusion limits and discovery sensitivities when available.  We note that for the Wino case the expected discovery sensitivity at MuC 3,10 is close to the kinematic limit of $\sqrt{s}/2$, hence the corresponding exclusion limits do not significantly extend the mass reach. With the exception of the FCC-hh~\cite{Saito:2019rtg}, being expected to cover pure wino scenarios up to about 6.5~TeV and pure higgsino up to 1.6~TeV, the MuC is one of the most promising proposed machines to cover this specific experimental signature.

\FloatBarrier
\section{Conclusions and outlook}
\label{sec:conclusions}
This work investigated the sensitivity of a future high-energy muon collider to new electroweak multiplets in compressed sub-GeV mass spectra exploiting the “disappearing” tracks signature.

A realistic simulation of the beam induced background was employed to prove for the first time the feasibility of this search even in the harsh environment predicted at the muon collider. We find that a combination of selections on the time and spatial correlation of pair of hits in neighbouring detector layers, followed by simple cuts on the reconstructed track quality, reduce the overwhelming BIB rate to manageable levels. We note that some of these BIB suppression techniques are relevant, or could be optimised, for other long-lived signatures at the muon collider, for instance displaced vertexes, displaced leptons, dark showers. Thus our result opens an important avenue to expand the physics programme of the muon collider into the realm of LLP.

As a byproduct of the BIB study, we have also derived the tracklet reconstruction efficiency based on the full detector simulation, mapping from the generator level quantities (decay radius, $\theta$) of short-lived charged particles, to the reconstructed tracklet, in a model-independent manner, and including a realistic \pT smearing. This efficiency map can be used to derive the expected reach of disappearing tracks at a muon collider for arbitrary models.

We note that the minimal reconstructible tracklet length, given by the position of the second double layer of the vertex detector is very close to the decay length predicted by pure higgsino models and urge future detector layouts not to adopt larger radii for these detector layers.

Using the results from the simulation, we have furthermore studied the sensitivity of a $\sqrt{s}=3$~TeV and $\sqrt{s}=10$ TeV muon collider with a simple cut-and-count strategy selecting events with an additional energetic photon radiation, taking as a motivation the well studied MSSM pure wino and pure higgsino models. When considering the mass-lifetime plane, we find that a 3 (10) TeV collider can cover masses up to 1.43~TeV (4.7)~TeV. Our analysis achieves discovery sensitivity in the lifetime region between 0.01 and 10~ns, narrowing down to 0.1-10~ns for large masses. When our results are phrased in terms of pure wino and higgsino (with proper lifetimes of 0.2~ns and 0.02~ns, respectively) we find that the thermal cases of 2.7 TeV and 1.1 TeV, respectively, can be probed at the 5-$\sigma$ level by a 10~TeV muon collider with ten ab$^{-1}$ of data is collected. This guarantees the discovery of thermal wino and higgsino DM, or else exclude the last standing bulwark of minimal WIMP dark matter models.

A high-energy muon collider has the potential to strongly extend the reach for high mass compressed states and has the potential to make a decisive statement on the phase space favoured by minimalistic SM extensions aimed at solving the dark matter problem, as well as other models featuring weak multiplets with masses above or about the TeV scale.

\acknowledgments
We are grateful to Beate Heinemann, Iacopo Vivarelli, Lawrence Lee, Simone Pagan Griso, Claudia Merlassino, Zhen Liu,  David Curtin, and Asimina Arvanitaki for useful discussions. We would like to thank Dario Buttazzo and Roberto Franceschini for a fruitful e-mail exchange that led to the update of Figures 12 and 13. We thank the International Muon Collider Collaboration for fostering this work.  The work of RC was supported in part by the Perimeter Institute for Theoretical Physics (PI), by the Canada Research Chair program, and by a Discovery Grant from the Natural Sciences and Engineering Research Council of Canada. Research at PI is supported in part by the Government of Canada through the Department of Innovation, Science and Economic Development Canada and by the Province of Ontario through the Ministry of Colleges and Universities. JZ is supported by the {\it Generalitat Valenciana} (Spain) through the {\it plan GenT} program (CIDEGENT/2019/068). This work has benefited from computing services provided by the German National Analysis Facility (NAF).

\FloatBarrier
\bibliographystyle{JHEP}  
\bibliography{references}  

\end{document}